\documentstyle[epsf]{article}

\newcommand{\caret}[1]{\hat{#1}}
\newcommand{\Tr}[1]{{\rm Tr}{#1}}

\begin{document}
\begin{flushleft}
{\bf SAGA-HE-137}
\end{flushleft}

\bigskip

\bigskip

\centerline{\bf Non-perturbative renormalization group equations}
\centerline{\bf and approximate discrete $\gamma_5$-symmetry}
\centerline{\bf in quantum hadrodynamics}

\bigskip

\bigskip

\centerline{\bf Hiroaki Kouno$^{**}$, Manabu Nakai, Akira Hasegawa}
\centerline{and}
\centerline{\bf Masahiro Nakano$^*$}

\centerline{Department of Physics, Saga University, Saga 840, Japan}
\centerline{$^{*}$University of Occupational and Environmental Health, Kitakyushu 807, Japan}

\centerline{** :e-mail address, kounoh@cc.saga-u.ac.jp}

\bigskip

\bigskip

\bigskip

\centerline{\bf Abstract}

\bigskip

\noindent
The quantum hadrodynamics are studied by using non-perturbative renormalization group equations. 
Approximate discrete $\gamma_5$-symmetry is studied. 
It is shown that the contributions of one-loop diagrams are important for effective potential. 
A relation between the local-potential approximation and the usual Hartree approximation is discussed. 
The local-potential approximation includes contribution of the Fock term to nucleon self-energy, contribution in the usual random phase approximation to meson self-energy and vertex corrections in part as well as the Hartree contribution to the effective potential. 
Approximate discrete $\gamma_5$-symmetry is studied phenomenologically in the Hartree approximation. 
This approximate symmetry may exist in nuclear matter.

\vfill\eject


\section{Introduction}

In the past two decades, nuclei and nuclear matter have been studied in the framework of quantum hadrodynamics (QHD).\cite{rf:Walecka,rf:Serot} 
The meson mean-field theory of nuclear matter\cite{rf:Walecka} has produced successful results to account for the saturation properties at normal nuclear density. 
Following these successes, many studies and modifications have been performed on relativistic nuclear models. 
One of these modifications is the inclusion of vacuum fluctuation effects, which cause divergences in physical quantities when they are naively calculated. 
Chin\cite{rf:Chin} estimated vacuum fluctuation effects in the Hartree approximation by using a renormalization procedure, and found that vacuum fluctuation effects make the incompressibility of nuclear matter smaller and closer to the empirical value than in the original Walecka model. 

Although the relation between QHD and the underlying fundamental theory, i.e., QCD, is an open question, it is natural that QHD is not valid at very high-energy. 
In this point of view, a cutoff or a form factor should be introduced into the theory of QHD. 
One may introduce the cutoff\cite{rf:Kohmura} or the form factor\cite{rf:Furn} to avoid the instability of the meson propagators in the random phase approximation (RPA)\cite{rf:Kurasawa}. 
Cohen\cite{rf:Cohen} introduced a four-dimensional cutoff into the relativistic Hartree calculation and found that the vacuum energy contribution may be somewhat different from that in the ordinary renormalization procedures, if the cutoff is not so large. 

One troublesome problem in use of the cutoff is that physical results depend on a value of the cutoff and a shape of the regulator which are introduced into the theory by hand. 
It is difficult to determine a suitable value of cutoff and a appropriate shape of the regulator phenomenologically. 
For the high-energy physics near the limitation of the theory of the particle physics, Lepage\cite{rf:Lepage} proposed to use the extended effective action (Lagrangian) to eliminate such a dependence on cutoff which is by hand introduced into the theory, rather than to search the phenomenologically favorable cutoff and regulator (or form factor). 
In the series of papers\cite{rf:Kouno1,rf:Kouno2,rf:Kouno3}, we have studied nuclear matter properties and the vertex corrections in the framework of the cutoff field theory, eliminating the cutoff dependence by using the Lepage's method. 

The extended action in that method is essentially the same as the one in the non-perturbative renormalization group (NPRG) equation. \cite{rf:Wilson,rf:Wegner,rf:Nicoll,rf:Polch,rf:Hasen,rf:Wetterich,rf:Clark,rf:Morris,rf:Aoki1,rf:Aoki2,rf:Aoki3} 
In fact, the former is the perturbative approximation for the later. 
Therefore, it is natural to use the NPRG equations in QHD (more generally in nuclear physics), since the QHD has strong couplings. 

In this paper, we study QHD using the NPRG equation. 
For simplicity, we restrict our discussion to the one at the zero-density. 
We also only treat the nucleon and the $\sigma$-meson in the NPRG equation. 
Extension to the finite baryon density and inclusion of the other meson are the problems in the furture. 
This paper is organized as follows. 
In {\S}2, we derive the NPRG equations for the $\sigma$-nucleon system. 
In {\S}3, we study the discrete $\gamma_5$-symmetry by solving the NPRG equations in the local-potential approximation.\cite{rf:Nicoll,rf:Hasen,rf:Clark,rf:Morris,rf:Aoki1,rf:Aoki2} 
It is shown that $approximate$ $\gamma_5$-symmetry may be preserved to some extent in analogy of the exact symmetry.\cite{rf:Clark} 
In {\S}4, the relation between the local-potential approximation and the traditional Hartree approximation in QHD is discussed. 
The local-potential approximation includes contribution of the Fock term to nucleon self-energy, contribution in the random phase approximation (RPA) to meson self-energy and vertex corrections in part as well as the Hartree contribution to the effective potential. 
In {\S}5, the approximate $\gamma_5$-symmetry is studied phenomenologically by using the Hartree approximation. 
It is shown that this approximate symmetry may exist in nuclear matter.
Section 6 is devoted to a summary. 

\section{Non-perturbative renormalization group equations \\in quantum hadrodynamics}

We use the NPRG equations with a sharp cutoff function $\theta (\Lambda -\vert p\vert )$\cite{rf:Wegner,rf:Nicoll,rf:Hasen,rf:Clark,rf:Morris,rf:Aoki1,rf:Aoki3} where $\Lambda$ is cutoff and $p$ is four momentum in Euclidean space. 
Using this cutoff function, the Euclidean action $S[\Theta, \Lambda]$ is given by 
\begin{eqnarray}
S[\Theta; \Lambda ]=\sum_n{{1}\over{n!}}
\int{{d^4p_1}\over{(2\pi )^4}}...\int{{d^4p_n}\over{(2\pi )^4}}
(2\pi )^4\delta^4(p_1+...+p_n)
\nonumber\\
\times\theta(\Lambda -\vert p_1\vert )...\theta(\Lambda -\vert p_n\vert )
g_{{i_1}...{i_n}}(p_1...p_n; \Lambda )\Theta_{i_1}(-p_1)...\Theta_{i_n}(-p_n), 
\label{eq:2}
\end{eqnarray}
where $\Theta_i \in\{\phi ,\psi ,\bar{\psi}\}$, $\phi$ and $\psi$ are the nucleon field and $\sigma$-meson field and $g$ denotes a coupling, respectively. 
The generating functional $Z[J]$ for the Schwinger functions (the Euclidean-space Green functions) is given by
\begin{equation}
Z[J] =\int[d\Theta]\exp\left(-S[\Theta ;\Lambda]+\int d^4xJ_i(x)\Theta_i(x)\right). 
\label{eq:Ad1} 
\end{equation}
where $J_i$ are sources to the field $\Theta_i$. 

Next we put $\Lambda =\Lambda_0\exp{(-t)}$ where $\Lambda_0$ is the initial value of the cutoff and $t$ is a dimensionless parameter which runs from 0 to infinity. 
The NPRG equations are derived by varying the cutoff and by considering the response of the theory subject to the constraint that the Schwinger functions remain unaltered. 
(For the details of the derivation, see Ref.~18). )
It is given by a differential equation with respect to $t$ as follows. 
\begin{eqnarray}
{{d}\over{dt}}\caret{S}[\caret{\Theta} ;t]&=&{{1}\over{2}}{{1}\over{\delta t}}\int_{1-\delta t\leq \vert \caret{p}\vert \leq 1}{{d^4\caret{p}}\over{(2\pi)^4}}{\rm str}\ln{\caret{S}_{ij} (\caret{p},-\caret{p})}
\nonumber\\
&-&{{1}\over{2}}{{1}\over{\delta t}}\int_{1-\delta t\leq \vert \caret{p}\vert ,\vert \caret{q}\vert \leq 1}{{d^4\caret{p}}\over{(2\pi )^4}}{{d^4\caret{q}}\over{(2\pi )^4}}(-)^{h_{i}}
{{\delta \caret{S}[\caret{\Theta} ;t ]}\over{\delta \caret{\Theta}_i(\caret{p})}}\caret{S}^{-1}_{ij}(\caret{p},\caret{q})
{{\delta \caret{S}[\caret{\Theta} ;t ]}\over{\delta \caret{\Theta}_j(\caret{q})}}
\nonumber\\
&+&\left[ 
4-\sum_i \int_{\vert \caret{p}\vert \leq 1}{{d^4\caret{p}}\over{(2\pi )^4}}\Theta_i(-\caret{p})
\left( d_{\Theta_{i}}-\gamma_{\Theta_{i}}+\caret{p}^\mu{{\partial}\over{\partial \caret{p}^\mu}}\right) {{\delta}\over{\delta \caret{\Theta}_i(\caret{p})}}
\right] 
\nonumber\\
&\times &S[\caret{\Theta} ;t ], 
\nonumber\\
\label{eq:3}
\end{eqnarray}
where $d_{\Theta_i}$ and $\gamma_{\Theta_i}$ are the naive and the anomalous dimension of field $\Theta_i(x)$, respectively, and 
$h_i$ is zero if $\Theta_i$ is a boson field and unity if $\Theta_i$ is a fermion field.  
The symbol "str" in equation (\ref{eq:3}) denotes supertrace.  
We remark that the four momentum and the field in Eq. (\ref{eq:3}) have been rescaled and become dimensionless by the replacement
\begin{equation}
\caret{p}^\mu ={{p^\mu}\over{\Lambda}},~~~~~\caret{\Theta}_i(\caret{p})=D_{\Theta_i}(\Lambda )\Theta_i(p), 
\label{eq:5}
\end{equation}
where $D_{\Theta_i}(\Lambda )$ is given by following equation
\begin{equation}
\Lambda{{d}\over{d\Lambda}}\ln{D_{\Theta_i}(\Lambda )}=4-d_{\Theta_i}+\gamma_{\Theta_i}. 
\label{eq:6}
\end{equation}
Correspondingly, the couplings have been rescaled and become dimensionless by the replacement 
\begin{equation}
\caret{g}_{i_{1}...i_{n}}(\caret{p}_{1},...,\caret{p}_{n};\Lambda) 
=\Lambda^{4n-4}g_{i_{1}...i_{n}}(p_1,...,p_n;\Lambda )
\prod_{j=1}^{n}D_{\Theta_{i_{j}}}^{-1}(\Lambda ). 
\label{eq:Ad2}
\end{equation}
Using these dimensionless quantities, $\caret{S}(\caret{\Theta} ;t)$ is defined by
\begin{equation}
\caret{S}(\caret{\Theta};t)=S(\Theta ;\Lambda )
\label{eq:Ad9}
\end{equation}
The inverse propagators $\caret{S}_{ij}(\caret{p},\caret{q})$ are given by 
\begin{equation}
\caret{S}_{ij}(\caret{p},\caret{q})=(-)^{H_{ij}}{{\delta^2 \caret{S}[\caret{\Theta} ; t]}\over{\delta \caret{\Theta}_j(\caret{q})\delta \caret{\Theta}_i(\caret{p})}},
\label{eq:4}
\end{equation}
where $H_{ij}$ is unity if both the field $\Theta_i$ is an even element of the Grassmann algebra and the field $\Theta_j$ is an odd element of the Grassmann algebra and vanishes otherwise. 

In this section, below(except for (\ref{eq:12})), we work only with dimensionless fields and momentum vectors and drop the caret notation on all quantities for the simplicity of the notation. 

Equation (\ref{eq:3}) is exact for all quantum corrections including any higher order correction in perturbation theory. 
However, it is the functional differential equation and difficult to be solved. 
To solve it, we ignore all interactions involving derivative couplings and furthermore set the wavefunction renormalizations to unity, therefore, the anomaous dimension $\gamma_{\Theta_i}$ is neglected. 
This is called the local-potential (or action ) approximation \cite{rf:Nicoll,rf:Hasen,rf:Clark,rf:Morris,rf:Aoki1,rf:Aoki2} and the lowest-term of a systematic momentum (or derivative) expansion. 
The locally approximated action is given by
\begin{equation}
S[\Theta ={\rm{const.}};\Lambda ] =(2\pi )^4\delta^4(0)U(\Theta; t), 
\label{eq:7}
\end{equation}
where a generalized potential is given by
\begin{equation}
U[\Theta ;t ] =\sum_nC_{{i_1}...{i_n}}(t)\Theta_{i_1}...\Theta_{i_n}, 
\label{eq:8}
\end{equation}
$\Theta_i$ is defined by
\begin{equation}
\Theta_i(p)=(2\pi )^4\delta^4(p)\Theta_i.   
\label{eq:9}
\end{equation}
and $C_{i_{1}...i_{n}}(t)$ denotes the coupling which does not depend on $p$. 
Below we do not denote the $t$-dependence of the couplings $C
_{i_{1}...i_{n}}$ explicitly for the simplicity of the notation. 

In this approximation, Eq. (\ref{eq:3}) is reduced to 
\begin{eqnarray}
{{\partial}\over{\partial t}}U(\Theta ,t)
={{1}\over{2}}{{1}\over{\delta t}}\int_{1-\delta  t\leq \vert p \vert \leq 1}{{d^4p}\over{(2\pi )^4}}{\rm str} \ln{[T_{ij}(p)+(-)^{h_{ij}}{{\partial^2}\over{\partial \Theta_j\partial\Theta_i}}U(\Theta , t)]}
\nonumber\\
+[4-\sum_{\Theta_{i}}d_{\Theta_i}\Theta_i{{\partial}\over{\partial\Theta_i}}]U(\Theta ,t). 
\label{eq:10}
\end{eqnarray}
$T_{ij}(p)$ in Eq. (\ref{eq:10}) is a kinetic term and given by
\begin{equation}
T(p)=
\left(
\begin{array}{ccc}
      p^2 & 0 & 0 \\
      0   & 0 & -i\gamma^T\cdot{ p}\otimes I_{i} \\
      0   & -i\gamma\cdot{ p}\otimes I_{i} & 0    
\end{array}
\right) ,
\label{eq:11}
\end{equation}
where $\gamma$ is the $\gamma$-matrix in Euclidean space, $I_{i}$ is the unit matrix in isospin space, and the symbol $\otimes$ denotes a direct product. 
We remark that, in the local-potential approximation, the second replacement in Eqs. (\ref{eq:5}) and the replacement (\ref{eq:Ad2}) are reduced to 
\begin{equation}
\caret{\Theta_i}
=\Lambda^{-d_{\Theta_i}}\Theta_i
~~~~~{\rm and}~~~~~
\caret{C}_{i_{1}...i_{n}} 
=\Lambda^{-d_{i_{1}...i_{n}}^c}
C_{i_{1}...i_{n}}, 
\label{eq:12}
\end{equation}
where ${\displaystyle d_{i_{1}...i_{n}}^c=4-\sum_{i=1}^nd_{\Theta_{i}}}$ 
is a naive dimension of the coupling $C_{i_{1}...i_{n}}$ without the caret. 
In Eq. (\ref{eq:10}), we see that the r. h. s. has a one-loop structure with "full propagators" which include the quantum corrections in higher energy region. 

Furthermore, if we assume the generalized potential $U(\Theta, t)$ depends only on the real scalar field $\phi$ and $\sigma =\bar{\psi}_i\psi_i$ (we remark that the sum is taken over both the spinor and the isospin indices in $\sigma$), Eq. (\ref{eq:10}) is reduced to 
\begin{eqnarray}
{{\partial}\over{\partial t}}U(\phi ,\sigma ;t)
=4U(\phi , \sigma ;t)-\phi U_{\phi}(\phi ,\sigma ;t)-3\sigma U_{\sigma}
(\phi, \sigma ;t)
\nonumber\\
+{{1}\over{16\pi^2}}\ln{(1+U_{\phi\phi}(\phi, \sigma ;t))}
-{{\lambda}\over{4\pi^2}}\ln{(1+U_\sigma^2(\phi ,\sigma ;t))}
\nonumber\\
+{{1}\over{16\pi^2}}\ln{(1+\Omega (\phi ,\sigma ;t))}, 
\label{eq:13}
\end{eqnarray}
where 
\begin{equation}
\Omega =2{{\sigma U_{\sigma}}\over{1+U_\sigma^2}}
[U_{\sigma\sigma}-U_{\sigma\phi}\left({{1}\over{1+U_{\phi\phi}}}\right) U_{\phi\sigma } ], 
\label{eq:14}
\end{equation}
$\lambda$ is the degree of freedom of isospin of fermion (in nucleon system, $\lambda=2$) and $U_{\sigma}={{\partial }\over{\partial \sigma}}U$ etc.. 
As is seen in {\S}3, the first, the second and the third terms in the right-hand side (r. h. s.) of Eq. (\ref{eq:13}) reflects the effects of the naive dimensions of the original couplings without the caret. 
Below, we call this part "naive dimensional part". 
The fourth and the fifth terms in the r. h. s. of Eq. (\ref{eq:13}) look like a bosonic and a fermionic one-loop contributions, respectively. 
Below, we call them "bosonic part" and "fermionic part", respectively. 
The last term in r. h. s. of Eq. (\ref{eq:13}) has both bosonic and fermionic contributions. 
We call this part "boson-fermion (BF) part". 
As is seen in (\ref{eq:13}), the BF part has at least one $\sigma$. 
Therefore, this term contributes $directly$ only to the evolutions of the couplings of the the interactions which have $at$ $least$ $one$ $\sigma$. 

Now, the functional differential equation (\ref{eq:3}) is reduced to the partial differential equation (\ref{eq:13}). 
In the next section, we study approximate discrete $\gamma_5$-symmetry using Eq. (\ref{eq:13}).

\section{Approximate discrete $\gamma_5$-symmetry in quantum hadrodynamics}

In this section, we use the coupling constant $C_{l,m}$ which is defined by
\begin{equation} 
U(\phi ,\sigma ;t)=\sum_{l,m=0}^{\infty}C_{l,m}\phi^l\sigma^m 
\label{eq:15}
\end{equation}
rather than ones defined in Eq. (\ref{eq:8}). 
In the actual calculation, we truncate infinite series in Eq. (\ref{eq:15}) as follows. 
\begin{equation}
U(\phi ,\sigma ;t)=\sum_{l=0}^{8}C_{l,0}\phi^l+\sum_{l=0}^{5}C_{l,1}\phi^l\sigma+\sum_{l=0}^{2}C_{l,2}\phi^l\sigma^2. 
\label{eq:16}
\end{equation}
This means that we neglect the higher order interactions $\phi^l\sigma^m$ whose naive dimensions are greater than 8. 
If we put (\ref{eq:16}) into the first three terms of the r. h. s. of Eq. (\ref{eq:13}), we get
\begin{eqnarray}
\sum_{l=0}^8(4-l)C_{l,0}\phi^l+\sum_{l=0}^5(1-l)C_{l,1}\phi^l\sigma+\sum_{l=0}^2(-2-l)C_{l,2}\phi^l\sigma^2. 
\nonumber\\
=\sum_{l=0}^8d^c_{l,0}C_{l,0}\phi^l+\sum_{l=0}^5d^c_{l,1}C_{l,1}\phi^l\sigma+\sum_{l=0}^2d^c_{l,2}C_{l,2}\phi^l\sigma^2. 
\label{eq:17}
\end{eqnarray}
We see that the naive dimension $d^c_{l,m}$ of the couplings $C_{l,m}$ appears before the corresponding coupling. 

In high energy physics, a discrete $\gamma_5$-symmetry is sometimes imposed on the generalized potential,\cite{rf:Clark} i.e, $U(\phi, \sigma ;t)$ is invariant under the transformation, 
\begin{equation}
\phi^\prime=-\phi,~~~\psi^\prime =\pm\gamma_5\psi,~~~~~\bar{\psi}^\prime=\mp\bar{\psi}\gamma_5,~~~\sigma^\prime =-\sigma, 
\label{eq:18}
\end{equation}
where the plus (minus) sign in the second equation and the minus (plus) sign in the third equation are for proton (neutron). 
If we extend the theory to include pion, this discrete symmetry may be embedded in the continuous chiral symmetry. 
The discrete $\gamma_5$-symmetry forbides the "odd" interaction terms with $l+m$ equal an odd integer. 
Therefore, explicit fermion mass terms are forbidden as in the chiral symmetric theory. 

However, in QHD, the nucleon already has a explicit mass term, so we do not impose this symmetry on $U(\phi, \sigma ;t)$. 
However, if we impose this symmetry on the initial conditions for the differential equation, i.e., put all of the "odd" coupling $C_{l,m}$ with $l+m$ equal an odd integer to zero at $t=0$, 
they are remains zero through the whole evolution. 
This results is understood as follow. 
If we expand the Eq. (\ref{eq:13}) in power of $\phi$ and $\sigma$, each term of the expansion has a form as 
\begin{equation}
\phi^l\sigma^m{{dC_{l,m}}\over{dt}}=\phi^l\sigma^m\sum_{i}A_i{{C_{l_{1},m_{1}}...C_{l_{n_{i}},m_{n_{i}}}}\over{(1+2C_{2,0})^{j_{i}}(1+C_{0,1}^2)^{k_{i}}}}
\label{eq:19}, 
\end{equation}
where $j_i$ and $k_i$ are nonnegative integers, and $A_i$ is a constant which is independent of the fields and couplings. 
The sum of the r. h. s. in Eq. (\ref{eq:19}) correspond to the variation of the coupling $C_{l,m}$ in the infinitesimal interval $\delta t$. 
It is easy to observe that the sum
\begin{equation}
\sum_{{s_i}=1}^{n_i}\left(l_{s_{i}}+m_{s_{i}}\right)
\label{eq:20}
\end{equation}
is an odd integer if $l+m$ is an odd integer and is an even integer if $l+m$ is an even integer.
This means that, in the case of the odd coupling evolutions equation, the numerators in the sum of Eq. (\ref{eq:19}) have an $odd$ number of $odd$ couplings. 
Therefore, the numerators have at least $one$ odd coupling. 
For example, a $\phi^3$-term in the expansion of the bosonic part 
\begin{equation}
{{1}\over{16\pi^2}}\ln{(1+U_{\phi\phi}(\phi, \sigma ;t))}
\label{eq:21}
\end{equation}
is 
\begin{equation}
{{1}\over{(3!)(16\pi^2)}}\left(120{{C_{5,0}}\over{1+2C_{2,0}}}-432{{C_{3,0}C_{4,0}}\over{(1+2C_{2,0})^2}}+432{{C_{3,0}^3}\over{(1+2C_{2,0})^3}}\right)\phi^3. 
\label{eq:22}
\end{equation}
The numerators in each term of Eq. (\ref{eq:22}) have at least one odd coupling, therefore Eq. (\ref{eq:22}) is zero at $t=0$ if all odd couplings are zero at $t=0$. 
The other coefficients of $\phi^3$ in the r. h. s. of (\ref{eq:13}) also become zero at $t=0$. 
Therefore, we get
\begin{equation}
{{dC_{3,0}}\over{dt}}=0  
\label{eq:23}
\end{equation}
at $t=0$. 
Samely, the other odd couplings are also unchanged. 
Therefore, the odd couplings remain zero through the whole evolution if all of the odd couplings are put to be zero at $t=0$. 
This mechanism may raise the possibility that approximate discrete $\gamma_5$-symmetry remains through the evolution as well as the exact symmetry. 
Below, we examine this possibility. 

In figs. \ref{fig:1}$\sim$\ref{fig:5}, we show some results with the initial conditions with $C_{2,0}=0.05$, $C_{4,0}=2.5$, $C_{0,1}=0.3$, $C_{1,1}=-2.0$ and the other couplings equal to zero at $t=0$. 
( We remark that we use the negative value of $C_{1,1}$ to correspond to the usual definition of Yukawa coupling in QHD.\cite{rf:Serot} See also Eq. (\ref{eq:24}) in {\S}4.) 
In this case, only the fermion mass term $C_{0,1}\sigma$ break the discrete $\gamma_5$-symmetry at the initial. 
Although, in this paper, we show only the results with one set of initial conditions, a qualitative features are similar if we change the initial conditions to some extent. 


It is seen that the couplings with the positive naive dimension rapidly increase in the large $t$ region, while the coupling with the negative naive dimension rapidly decrease in the same region of $t$. 
The couplings whose naive dimensions are zero seem to approach a constant value in the large $t$ limit. 
Behavior of the evolutions of the couplings at large $t$ reflects the effects of the naive dimensions $d_{l,m}^c$ of the couplings which appear the naive dimensional part (\ref{eq:17}). 
These couplings evolute approximately proportional to $\exp{(d^c_{l,m}t)}$ at large $t$, e.g., $C_{2,0}\sim \exp{(2t)}$, $C_{4,0}\sim$const., $C_{6,0}\sim\exp{(-2t)}$. 
However, we remark that, as is seen in Fig. \ref{fig:3}, 
the couplings with a negative dimension may increase in the small $t$ region, by the effects of the complicated interactions included in the bosonic, the fermionic and BF parts in the r. h. s. of Eq. (\ref{eq:13}). 
This means that these couplings may be important if we interested in the "high energy" physics whose energy scale is close to $\Lambda_0$. \cite{rf:Lepage,rf:Clark,rf:Kouno1,rf:Kouno2,rf:Kouno3} 

As is seen in {\S}2, the BF part does not contribute directly to the evolution of the coupling $C_{l,0}$. 
In Figs. \ref{fig:6} and \ref{fig:7}, we show the contributions of the bosonic and the fermionic parts in the r. h. s. of Eq. (\ref{eq:13}) for the evolution of $C_{4,0}$ coupling. 
The initial conditions are the same as in Figs. \ref{fig:1}$\sim$\ref{fig:5}. 
The bosonic part and fermionic parts cancel each other only partially, and they yield some amount of the total contribution. (See Fig. \ref{fig:8}.)
In Figs. \ref{fig:9}$\sim$\ref{fig:11}, we show the contributions of the bosonic, the fermionic and the BF parts for the evolution of $C_{1,1}$ coupling. 
The initial conditions are the same as above. 
From the figures, we see that the contributions of the bosonic and the fermionic parts hardly cancel the dominant BF part. 
However, the whole contribution is much smaller than in the case of $C_{4,0}$. 
(See Fig. \ref{fig:12}.) 

To see the approximate discrete $\gamma_5$-symmetry, in Fig. \ref{fig:13}, we show the absolute value of the couplings $C_{l,0}$ at $t=4$ where almost all quantum corrections have been included into the effective generalized potential. 
In this figure, we have dividedized the coupling by the factor 
\begin{equation}
(Q^\prime )^{4-l}g_s^l=
(Q^\prime )^{d_{l,0}^c}g_s^l=\left( {{Q\exp{(t)}}\over{\Lambda_0}}\right) ^{d^c_{l,0}}g_s^l, 
\label{eq:Ad81201}
\end{equation}
where $g_s=-C_{1,1}$ and $Q$ is a typical energy scale of the system, due to the following two reasons. 
The first, the coupling $C_{l,0}$ is approximately proportional to the factor $\exp{(d^c_{l,0}t)}=\exp{\{ (4-l)t\}}$ as is seen above. 
Therefore, we have divided the couplings by $(Q^\prime )^{d^C_{l,0}}$ which is proportional to $\exp{\{ (4-l)t\}}$ to eliminate the exponential $t$-dependence of the couplings. 
The second, the evolution of the coupling $C_{l,0}$ is the order $C_{1,1}^l$ at each step of the the NPRG equation, since the equation has the one-loop structure. 
For example, $C_{3,0}$ is the order of $(C_{1,1})^3$ as is seen in Fig. \ref{fig:14}. 
Therefore, we have divided the couplings by $\vert C_{1,1}^l\vert =g_s^l$. 
Note, this normalization is equivalent to normalized the original couplings without the caret by the factor $Q^{4-l}g_s^l$. 
In calculations, we put $Q/\Lambda_0=0.5$. 
In Fig. \ref{fig:13}, we see that the odd couplings are somewhat suppressed in comparison with the even couplings in the neighorhood, except for the coupling $C_{1,0}$ which is not small. 

It seem that approximate $\gamma_5$-symmetry is hold to some extent through the evolution except for $C_{1,0}$. 
However, nonvanishing value of the coupling $C_{1,0}$ induces the important effects. 
In fig. \ref{fig:15}, we show the generalized potential $U$ at $\sigma =0$ when $t$ becomes 4. 
It is seen that $U$ does not have minimum at $\phi =0 $ because of the nonvanishing $C_{1,0}$. 
The $\sigma$-meson field $\phi$ has non-zero vacuum expectation value $<\phi >$. 
Therefore, we must shift the $\phi$-field to the minimum to consider the real particle mass of the $\sigma$-meson. 
This shift may enlarge the other odd couplings as in the case of the spontaneous symmetry breaking (SSB) and make the nucleon mass larger, since $C_{1,1}<\phi>=-g_s<\phi >$ is positive. 
However, we remark that this enlargement of the odd couplings $C_{l,0}~(l\geq 3)$ is caused by the nonvanishing $C_{1,0}$ and not by the negative $C_{2,0}$ as in the case of SSB. 
This enhancement of the symmetry breaking may yield the possibility for the following scenario.

(1) In the initial bare action, there exists approximate $\gamma_5$-symmetry and the nucleon mass is not so large at $t=0$. 

(2) Quantum corrections yield the large nonvanishing $C_{1,0}$ and the field $\phi$ has a non-zero vacuum expectation value which causes the redefinition of $\phi$. 
This shift of field $\phi$ enlarge the other odd couplings and yields the large mass of nucleon. 

In {\S}5, we examine this scenario phenomenologically using the Hartree approximation. 


\section{Relation between the local-potential \\
approximation and the Hartree approximation}

In this section, we discuss the relations between the local-potential approximation in the NPRG equations and the traditional Hartree approximation in quantum hadrodynamics. \cite{rf:Walecka,rf:Serot,rf:Chin,rf:Cohen,rf:Kouno1,rf:Kouno2} 
We truncate the r. h. s. of Eq. (\ref{eq:13}), according to the observations obtained by solving the NPRG equations. 
First, we perform the inverse replacement of Eqs. (\ref{eq:5}) and (\ref{eq:Ad2}) (see also (\ref{eq:12})), i.e., perform the replacement
\begin{equation}
\Lambda \caret{p}^\mu=p^\mu,~~~~~\Lambda\caret{\phi}=\phi~~~~~\Lambda^3\caret{\sigma}=\sigma~~~~~\Lambda^{d_{l,m}^c}\caret{C}_{l,m}= C_{l,m} 
\label{eq:25}
\end{equation}
where we restored the caret notation defined in the beginning of the previous section. 
By these replacements, the naive dimensional part in Eq. (\ref{eq:13}) vanishes and we get 
\begin{eqnarray}
-\Lambda^{-3}{{\partial}\over{\partial \Lambda}}V(\phi ,\sigma ;\Lambda)
&=&{{1}\over{16\pi^2}}\ln{\{(p^2+V_{\phi\phi}(\phi ;\Lambda ))/\mu^2 \}}
\nonumber\\
&-&{{\lambda}\over{4\pi^2}}\ln{\{(p^2+(V_\sigma (\phi ,\sigma ;\Lambda ))^2)/\mu^2\}}
\nonumber\\
&+&{{1}\over{16\pi^2}}\ln{(1+\tilde{\Omega} (\phi ,\sigma ;\Lambda ))}+f(\Lambda , \mu ), 
\label{eq:26}
\end{eqnarray}
where the effective potential $V$ is defined by
\begin{equation}
V(\phi ,\sigma ;\Lambda ) = \Lambda^4U(\caret{\phi}, \caret{\sigma} ;t)=\sum_{l,m=0}C_{l,m}\phi^l\sigma^m. 
\label{eq:27}
\end{equation}
and 
\begin{equation}
\tilde{\Omega}(\phi, \sigma ;\Lambda )=\Omega (\caret{\phi} ,\caret{\sigma} ;t)=2{{\sigma V_{\sigma}}\over{p^2+V_\sigma^2}}
\left[ 
V_{\sigma\sigma}-V_{\sigma\phi}\left({{1}\over{p^2+V_{\phi\phi}}}\right) V_{\phi\sigma } 
\right] . 
\label{eq:Ad6}
\end{equation}
We remark that the fields and the couplings without the caret in these equations are the ones after the replacements ({\ref{eq:25}) and have the ordinary dimensions. 
The $f(\Lambda ,\mu )$ in Eq. (\ref{eq:26}) is given by
\begin{equation}
f(\Lambda ,\mu )={{1}\over{16\pi^2}}\ln{(\mu^2/\Lambda^2)}
-{{\lambda}\over{4\pi^2}}\ln{(\mu^2/\Lambda^2)}
\end{equation}
where $\mu$ is an arbitrary scale parameter with the dimension of mass. 
The $f(\Lambda, \mu )$ does not depend on the field $\phi$ and $\sigma$. 
Therefore, this constant term contributes only to the evolution of the coupling $C_{0,0}$ which is the absolute value of the generalized potential at the origin $(\phi ,\sigma )=(0,0)$. 
Since $C_{0,0}$ must be determined phenomenologically in the actual calculations, we drop $f(\Lambda, \mu )$ below for simplicity of the notation. 

Next, we divide $V$ in three parts. 
\begin{equation}
V =  V^{(0)}(\phi;\Lambda )+V^{(1)}(\phi; \Lambda )\sigma+V_{higher~\sigma}(\phi ,\sigma; \Lambda), 
\label{eq:Ad8101}
\end{equation}
where 
\begin{eqnarray}
V^{(0)}(\phi;\Lambda )&=&\sum_{l=0}C_{l,0}\phi^l~~~~~V^{(1)}(\phi ;\Lambda )=\sum_{l=0}C_{l,1}\phi^l,~~~
\nonumber\\
&&{\rm and}~~~~~V_{higher \sigma}(\phi ,\sigma ;\Lambda )=\sum_{m=2}\sum_{l=0}C_{l,m}\phi^l\sigma^m. 
\label{eq:28}
\end{eqnarray}

In Fig. \ref{fig:16}, we show the $\sigma$-dependence of the effective potential $V$ at $\phi /\Lambda_0 =0.1$ and $t=4$. 
The initial conditions are the same as in the previous section. 
It is seen that $V$ is well approximated by the linear function of $\sigma$. 
Therefore, below, we neglect the higher $\sigma$ terms $U_{higher~\sigma}$. 
By this approximation, Eq. (\ref{eq:26}) is reduced to following two equations. 
\begin{eqnarray}
-\Lambda^{-3}{{\partial}\over{\partial \Lambda}}V^{(0)}(\phi ;\Lambda)
& = & {{1}\over{16\pi^2}}\ln{\{(p^2+V_{\phi\phi}^{(0)}(\phi ;\Lambda ))/\mu^2\}}
\nonumber\\
 & - & {{\lambda}\over{4\pi^2}}\ln{\{(p^2+(V^{(1)}(\phi ;\Lambda ))^2)/\mu^2\}}
\label{eq:29}\\
-\Lambda^{-3}{{\partial}\over{\partial \Lambda}}V^{(1)}(\phi  ;\Lambda)
 & = &  {{1}\over{16\pi^2}}V_{\phi\phi}^{(1)}(\phi ;\Lambda ){1\over{p^2+V_{\phi\phi}^{(0)}(\phi ;\Lambda )}}
\nonumber\\
&-& {{1}\over{8\pi^2}}V^{(1)}(\phi ; \Lambda )(V_{\phi}^{(1)}(\phi ,\Lambda ))^2\nonumber\\
&\times& {1\over{\{ p^2+V^{(0)}_{\phi\phi}(\phi ; \Lambda )\}\{ p^2+(V^{(1)}(\phi ;\Lambda ))^2\}}}. 
\label{eq:30}
\end{eqnarray}
We remark that the BF part of Eq. (\ref{eq:13}) dropped in the Eq. (\ref{eq:29}) as is seen in the {\S}2. 

Equations (\ref{eq:29}) and (\ref{eq:30}) are coupled differential equations for the potential $V^{(0)}(\phi ,\Lambda )$ and $V^{(1)}(\phi ,\Lambda )$. 
However, these differential equations are difficult to be integrated out since the unknown $\Lambda$-dependence of the couplings appears in the r. h. s. of the equations. 
Therefore, we divide the integrations into two steps. 
First, using the original renormalization group equation (\ref{eq:13}), we calculate the coupling evolutions from $t=0$ to $t=t^*=\ln{(\Lambda_0/\Lambda^*)}$ below which the couplings $C_{l,m}$ vary slowly. 
In figs. \ref{fig:17} and \ref{fig:18}, we show the $t$-dependence of the couplings $C_{2,0}/\Lambda_0^2$ and $C_{6,0}\Lambda_0^2$ after the replacement of (\ref{eq:25}). 
In the large region of $t$, in fact, they vary slowly. 
Therefore, below $t^*$, we perform the integration by putting the couplings in the r. h. s. in the differential equations to be constants. \cite{rf:Clark} 

For the constant couplings, we use the couplings at $\Lambda =0$ rather than ones at $\Lambda =\Lambda^*$. 
This approximation may be justified by the following reason. 
For example, consider the integral 
\begin{equation}
\int_0^{\Lambda^*}d\Lambda \Lambda^3
\ln{\{(p^2+V^{(0)}_{\phi\phi}(\phi ;\Lambda ))/\mu^2\}}. 
\label{eq:33}
\end{equation}
We expand this integral in powers of $\phi$ around $\phi =0$. 
\begin{eqnarray}
&~& \sum_{n=0} {1\over{n!}}{{\partial^n}\over{\partial \phi^n}}\int_0^{\Lambda^*}d\Lambda \Lambda^3
\ln{\{(p^2+V^{(0)}_{\phi\phi}(\phi ;\Lambda ))/\mu^2\}}\vert_{\phi =0}\phi^n. \nonumber\\
& = & \int_0^{\Lambda^*}d\Lambda \Lambda^3
\ln{\{(p^2+V^{(0)}_{\phi\phi}(\phi ;\Lambda ))/\mu^2\}}\vert_{\phi =0}
\nonumber\\
 & + & \sum_{n=1} {1\over{n!}}{{\partial^{n-1}}\over{\partial \phi^{n-1}}}\int_0^{\Lambda^*}d\Lambda \Lambda^3
{1\over{p^2+V^{(0)}_{\phi\phi}(\phi ;\Lambda )}}\vert_{\phi =0}\phi^n.
\label{eq:34}
\end{eqnarray}
In actual calculations, we must determine some lower terms of $\phi$ phenomenologically. 
For example, in the ordinary renormalization procedure,\cite{rf:Chin,rf:Serot} we must determine the coefficient of $\phi^n$ when $n$ is smaller than 5. 
In the case of the finite cutoff field theory, we may need to determine some higher terms to remove the cutoff-dependence of the physical results.\cite{rf:Lepage,rf:Kouno1,rf:Kouno2,rf:Kouno3} 
Therefore, the quantities we can calculate are the coefficients of the higher terms in Eq. (\ref{eq:34}). 
In such a term, the integrand has large power of $p$ in the denominator. 
This means that such a integral is dominated by the low-energy behavior of the integrand. 
Therefore, we use the low-energy couplings at $\Lambda =0$ rather than the high-energy couplings at $\Lambda =\Lambda^*$. 

Using the couplings at $\Lambda =0$, we get
\begin{eqnarray}
V^{(0)}(\phi ;0 ) & = & V^{(0)}(\phi ; \Lambda^*)
  +  
{{1}\over{16\pi^2}}\int_0^{\Lambda^*}d\Lambda\Lambda^3
\ln{\{(p^2+V_{\phi\phi}^{(0)}(\phi ;0 ))/\mu^2\}}
\nonumber\\
 & - & {{\lambda}\over{4\pi^2}}\int_0^{\Lambda^*}d\Lambda\Lambda^3 
\ln{\{(p^2+(V^{(1)}(\phi ;0 ))^2)/\mu^2\}}
\nonumber\\
 & = & V^{(0)}(\phi ; \Lambda^* )
 +  {1\over{2}}\int{{d^4p}\over{(2\pi )^4}}
\theta (\Lambda^* -\vert p \vert )
\ln{\{(p^2+V^{(0)}_{\phi\phi}(\phi ;0 ))/\mu^2\}}
\nonumber\\
 & - & {1\over{2}} \int{{d^4p}\over{(2\pi )^4}}
\theta (\Lambda^*- \vert p \vert )
\Tr{[I_s\otimes I_i]}
\ln{\{ (p^2+(V^{(1)}(\phi ; 0))^2)/\mu^2\}}
\nonumber\\
\label{eq:35}\\
V^{(1)}(\phi ;0 )
 & = & V^{(1)}(\phi ; \Lambda^* )+{{1}\over{16\pi^2}}V^{(1)}_{\phi\phi}(\phi ;0)\int_{0}^{\Lambda^*}d\Lambda\Lambda^3{1\over{p^2+V^{(0)}_{\phi\phi}(\phi ;0 )}}
\nonumber\\
 &- & {{1}\over{8\pi^2}}V^{(1)}(\phi , 0)(V_{\phi}^{(1)} (\phi ,0 ))^2
\int_0^{\Lambda^*}d\Lambda\Lambda^3
\nonumber\\
&\times &{1\over{\{ p^2+V^{(0)}_{\phi\phi}(\phi ; 0))\}\{p^2+(V^{(1)}(\phi ;0 ))^2\}}}. 
\nonumber\\
 & = & V^{(1)}(\phi ; \Lambda^* )+{{1}\over{2}}(-iV^{(1)}_{\phi\phi}(\phi ;0))
\int{{(-1)d^4p}\over{(2\pi )^4}}\theta(\Lambda^*-\vert p\vert )
i\Delta (p)
\nonumber\\
 & + & (-iV_{\phi}^{(1)}(\phi ,0 ))^2
\int{{(-1)d^4p}\over{(2\pi )^4}}
\theta (\Lambda^*- \vert p \vert )
i\Delta (p){{Tr{[iG(p)]}}\over{4\lambda}}, 
\nonumber\\
\label{eq:36}
\end{eqnarray}
where 
\begin{equation}
\Delta(p)={{-1}\over{p^2+V^{(0)}_{\phi\phi}(\phi ; 0)}}
\label{eq:Ad7}
\end{equation}
and 
\begin{equation}
G(p)={{i\gamma\cdot{p}-V^{(1)}(\phi ;0)I_s}\over{p^2+(V^{(1)}(\phi ;0))^2}}\otimes I_i
\label{eq:Ad8}
\end{equation}
are the $\sigma$-meson and the nucleon full propagators, respectively, $I_s$ is an unit matrix in spinor space, and the traces are taken over both the spinor and isospin indices. 

The first terms in the r. h. s. of Eqs. (\ref{eq:35}) and (\ref{eq:36}) play a role of counter-terms in the ordinary renormalization procedure. 
Second and third terms of the r. h. s. of Eq. (\ref{eq:35}) are the bosonic and the fermionic one-loop contributions, respectively. 
Equation (\ref{eq:36}) is an equation for calculating a scalar part of the nucleon self-energy, i.e., an effective nucleon mass. 
Especially, the third term in the r. h. s. corresponds to the Fock-term. 
In Figs. \ref{fig:19}(a) and (b), we show the graphical expressions of each terms in Eqs. (\ref{eq:35})  and (\ref{eq:36}). 


If we consider the second derivative of the both sides of Eq. (\ref{eq:35}) with respect to $\phi$, we get the following equation for calculating the effective $\sigma$-meson mass which is defined at the point where external momentum vanish. 
\begin{eqnarray}
V^{(0)}_{\phi\phi}(\phi ;0 ) & = & V^{(0)}_{\phi\phi}(\phi ;\Lambda^*)
\nonumber\\
 & + & 
{{1}\over{2}}(-iV^{(0)}_{\phi\phi\phi\phi}(\phi ;0)
\int{{(-1)d^4p}\over{(2\pi )^4}}
\theta (\Lambda^* -\vert p\vert )
i\Delta (p)
\nonumber\\
 &  + & 
{{1}\over{2}}(-iV^{(0)}_{\phi\phi\phi}(\phi ;0))^2
\int{{(-1)d^4p}\over{(2\pi )^4}}
\theta (\Lambda^* -\vert p\vert )
(i\Delta (p))^2
\nonumber\\
 & + & (-iV^{(1)}_{\phi\phi}(\phi ;0))
\int{{(-1)d^4p}\over{(2\pi )^4}}
\theta (\Lambda^* -\vert p\vert )
\Tr{\{iG(p)\}}(-1)
\nonumber\\
 & + & (-iV^{(1)}_{\phi}(\phi ;0))^2
\int{{(-1)d^4p}\over{(2\pi )^4}}
\theta (\Lambda^* -\vert p\vert )
\Tr{\{iG(p)iG(p)\}}(-1)
\nonumber\\
\label{eq:37}
\end{eqnarray}
The graphical expression of each term in Eq. (\ref{eq:37}) is shown in Fig. \ref{fig:20}(a). 

Similarly, 
If we differentiate the both sides of Eq. (\ref{eq:34}) with respect to $\phi$, we get the following equation for calculating effective Yukawa coupling with vanishing external momentum. 
\begin{eqnarray}
V^{(1)}_{\phi}(\phi ;0 )
 & = & V^{(1)}(\phi ; \Lambda^* )
+{{1}\over{2}}(-iV^{(1)}_{\phi\phi\phi}(\phi ,0))
\int{{(-1)d^4p}\over{(2\pi )^4}}
\theta (\Lambda^* -\vert p\vert )
i\Delta (p)
\nonumber\\
 & + & {{1}\over{2}}(-iV^{(1)}_{\phi\phi}(\phi ;0))(-iV^{(0)}_{\phi\phi\phi}(\phi ;0))
\int{{(-1)d^4p}\over{(2\pi )^4}}
\theta (\Lambda^* -\vert p\vert )
(i\Delta (p))^2
\nonumber\\
  &+&  2(-iV^{(1)}_{\phi}(\phi , 0))(-iV^{(1)}_{\phi\phi}(\phi ;0 ))
\int{{(-1)d^4p}\over{(2\pi )^4}}
\nonumber\\
&\times &\theta (\Lambda^* -\vert p\vert )
i\Delta (p){{\Tr{\{iG(p)\}}}\over{4\lambda}}
\nonumber\\
 & + & (-iV^{(1)}_{\phi}(\phi , 0))^2(-iV^{(0)}_{\phi\phi\phi}(\phi ;0 ))
\int{{(-1)d^4p}\over{(2\pi )^4}}
\nonumber\\
&\times &\theta (\Lambda^* -\vert p\vert )
(i\Delta (p))^2{{\Tr{\{iG(p)\}}}\over{4\lambda}}
\nonumber\\
 & + & (-iV^{(1)}_{\phi}(\phi ;0 ))^3
\int{{(-1)d^4p}\over{(2\pi )^4}}
\theta (\Lambda^* -\vert p\vert )
i\Delta (p){{\Tr{\{iG(p)iG(p)\}}}\over{4\lambda}}
\nonumber\\
\label{eq:38}
\end{eqnarray}
The graphical expression of each term in Eq. (\ref{eq:38}) is shown in Fig. \ref{fig:20}(b). 
Using the Eq. (\ref{eq:38}), we can calculate the effective Yukawa couplings. 

In fig. \ref{fig:21}, we show the $\phi$-dependence of $V^{(1)}$ which is obtained at $t=4$ by solving the NPRG equations. 
The initial conditions are the same as above. 
We see that $V^{(1)}$ can be well approximated by a linear approximation
\begin{equation}
V^{(1)}(\phi ;\Lambda )=C_{0,1}+C_{1,1}\phi =M^\prime -g_s\phi, 
\label{eq:24} 
\end{equation}
where $M^\prime =C_{0,1}$. 

If we put Eq. (\ref{eq:24}) into Eqs. (\ref{eq:35}) and regard $V_{\phi\phi}(\phi ;0)$ as a constant parameter in the r. h. s. in Eq. (\ref{eq:35}), 
we get the usual Hartree approximation.\cite{rf:Chin,rf:Serot,rf:Cohen,rf:Kouno1,rf:Kouno2}, since the bosonic loop in Eq. (\ref{eq:35}) yields only the constant contribution which is determined phenomenologically and it can be dropped. 
Under the same approximation, the second term of r. h. s. of Eq. (\ref{eq:36}) is dropped, Eq. (\ref{eq:36}) reduces to the equation for calculating the Fock-term contribution \cite{rf:Furn2} with vanishing external momentum, using the Hartree propagator perturbatively. 
Similarly, only the first and last terms of r. h. s. of Eqs. (\ref{eq:37}) and (\ref{eq:38}) remain, and we get the equation in the usual RPA\cite{rf:Kurasawa,rf:Furn,rf:Kohmura} and vertex correction\cite{rf:Kouno3}, respectively. 

The results presented here is the ones at zero baryon density. 
To justify the Hartree approximation at finite baryon density, it is needed to extend the NPRG equations to finite density. 
Furthermore, the effects of the other mesons, especially of the $\omega$-meson should be examined to study the nuclear matter. 
However, below, we assume the validity of the Hartree approximation at finite density in the $\sigma$-$\omega$ model. 
In fact, according to the calculations based on the Nuclear Schwinger-Dyson formalism,\cite{rf:Nakano1,rf:Nakano2,rf:Nakano3,rf:Hasegawa1,rf:Hasegawa2,rf:Mitsumori} the higher order corrections beyond the Hartree approximation seem to cancel each other.\cite{rf:Noda}  
In {\S}5, we examine the approximate $\gamma_5$-symmetry in the Hartree calculations in the $\sigma$-$\omega$ model at finite baryon density.

\section{Approximate $\gamma_5$-symmetry in the Hartree 
\\calculations in nuclear matter}

In this section, we examine an approximate $\gamma_5$-symmetry in $\sigma$-$\omega$ model \cite{rf:Walecka,rf:Chin,rf:Serot} in the framework of the finite cutoff field theory,\cite{rf:Cohen,rf:Lepage,rf:Kouno1,rf:Kouno2,rf:Kouno3} using the Hartree approximation. 
In this model,\cite{rf:Kouno1,rf:Kouno2} 
the energy density of nuclear matter is given by
\begin{eqnarray}
\varepsilon(\rho ,\Phi ,\Lambda^\prime )
 &= & V(\Phi )+V_{vacuum}(\Phi ; \Lambda^\prime )
-i\int{{d^4p}\over{(2\pi )^4}}\Tr{[G_{D}(p)]}
+{{C_v^2}\over{2M^2 }}\rho^2; 
\nonumber\\
\label{eq:39}
\\
V(\Phi ) & = & \sum_{l=0}^6{{c_l}\over{g_s^l}}\Phi^l;
\label{eq:Ad81101}
\\
V_{vacuum}(\Phi ; \Lambda^\prime ) &= & -{1\over{2}} \int{{d^4p}\over{(2\pi )^4}}
\theta (\Lambda^\prime- \vert p \vert )
\Tr{[I_s\otimes I_i]}
\ln{\{ (p^2+M^{*2})/\mu^2\}}
\label{eq:Ad81102}
\end{eqnarray}
In these equations, $\Lambda^\prime$ is a convenient cutoff which is introduced by hand for calculations and corresponds to $\Lambda^*$ in the previous section. 
$M$ is a nucleon mass, and $\Phi$, $c_l$, $M^*$ and $C_v$ are defined by
\begin{equation}
\Phi =g<\phi >,~~~~~c_l=C_{l,0}~~~~~M^*=M-\Phi,~~~~~{\rm and}~~~~~C_v={{g_vM}\over{m_v}}
\label{eq:40}
\end{equation}
, respectively, where $m_v$ and $g_v$ are $\omega$-meson mass and $\omega$-nucleon coupling, respectively, and $<\phi >$ is the ground state expectation value of $\phi$ at each baryon density. 
The baryon density $\rho$ is given by
\begin{equation}
\rho ={{\lambda}\over{2\pi^2}}p_F^3
\label{eq:41}
\end{equation}
where $p_F$ is a Fermi momentum. 
$G_{D}(p)$ is the density part\cite{rf:Walecka,rf:Chin,rf:Serot} of the nucleon propagator which depends $p_F$ explicitly, and is given by
\begin{eqnarray}
G_D(p)&=&(-i\gamma\cdot p +M^*I_s)2\pi\delta (p^{*2}+M^{*2})\theta (p_0)\theta (p_F-\vert {\bf p}\vert )\otimes I_i;
\nonumber\\
p^*_\mu&=&(p_4+iW_0,{\bf p});~~~~~p_0=ip_4, 
\label{eq:43}
\end{eqnarray}
where ${\bf p}$ are the three dimensional momentum and $W_0$ is given by
\begin{equation}
W_0=g_v<w_0>={{C_v^2}\over{M^2}}\rho 
\label{eq:80501}
\end{equation}
with  the ground-state expectation value of the time component $w_0$ of the $\omega$-meson field. 
$\Phi$ is determined by the equation of motion of the $\sigma$-meson, i.e., 
\begin{equation}
{{\partial }\over{\partial \Phi}}\varepsilon (\rho ,\Phi ;\Lambda^\prime )=0. 
\label{eq:Ad20}
\end{equation}

The first and the second terms in the r. h. s. of (\ref{eq:39}) correspond to the first and the third terms in the r. h. s. of (\ref{eq:35}). 
The first term plays a role of the "counter-term" in the ordinary renormalization procedure. 
We have truncated the higher-terms $\Phi^l$ ($l>6$) in it. 
Coupling $c_l$ is determined by 
the condition 
\begin{equation}
{{1}\over{l!}}{{\partial^l}\over{\partial \Phi^l}}\varepsilon (\rho =0, \Phi ;\Lambda^*)\vert_{\Phi =0}={{c^*_{l}}\over{g_s^l}}
\label{eq:44}
\end{equation}
where $c^*_l$  are effective or "physical" couplings which are determined phenomenologically. 
We choose 
\begin{equation}
{{c^*_l}\over{g_s^l}}=0~~~(l=0,1,3,4),~~~~~{\rm and}~~~~~{{c^*_2}\over{g_s^2}}={{M^2}\over{2C_s^2}}={{m_s^2}\over{g_s^2}}, 
\label{eq:45}
\end{equation}
where $C_s=g_sM/m_s$ and $m_s$ is a $\sigma$-meson mass. 
Note that the model depends on $m_s$, $m_v$, $g_s$, $g_v$ and $c_l$ only through the raitos $g_s/m_s$, $g_v/m_s$ and $c_l/g_s$. 
If  we use the nucleon mass $M=939$MeV, only four variable parameters remains, i.e., $C_s$, $C_v$, $c_5/g_s^5$ and $c_6/g_s^6$. 
We choose the four parameters to satisfy the saturation and compressional properties in nuclear matter.\cite{rf:Kouno2} 
For the saturation properties, we use the binding energy 15.75MeV at the saturation density $\rho_{st}=0.15$fm$^{-3}$. 
For the compressional properties, we use the results analized by Pearson.\cite{rf:Pearson} 
In table I, some available parameter sets are shown. 
We remark that the phenomenological determinations of $c_5$ and $c_6$ make the $\Lambda^\prime$-dependence of the physical results very small. \cite{rf:Kouno2}

First, for each parameter set, we expand the energy density in powers of $\Phi$ around $\Phi =0$ ( or $M^*=M$ )at zero baryon density as follows. 
\begin{equation}
\varepsilon (\rho =0,\Phi ;\Lambda^\prime )=\sum_{l=0}{1\over{l!}}{{\partial^l }\over{\partial \Phi^l}}\varepsilon (\rho=0,\Phi ;\Lambda^*)\vert_{\Phi =0}\Phi^l =\sum_{l=0}{{c_l^*}\over{g_s^l}}\Phi^l
\label{eq:46}
\end{equation}
In Figs. \ref{fig:22} and \ref{fig:23}, we show the absolute values of the couplings $c_l^*M^{l-4}/g_s^l~(1\leq l\leq 6)$ which are determined phenomenologically in this model as is seen above. 
We use the nucleon mass $M$ as the typical energy mass scale $Q$ which is used to normalized the couplings. 
In each case of the parameter set (PS), it seems that there are no explicit suppression of the odd couplings.

Next, we expand the energy density around $M^*=M^*_{st}$ at the saturation density $\rho_{st}$, i.e., 
\begin{equation}
\varepsilon (\rho =\rho_{st},\Phi ;\Lambda^\prime )=\sum_{l=0}{1\over{l!}}{{\partial^l }\over{\partial \Phi^l}}\varepsilon (\rho=\rho_{st},\Phi ;\Lambda^\prime )\vert_{\Phi =\Phi_{st}}\Phi^l =\sum_{l=0}{{b_l^*}\over{g_s^l}} \Phi^l, 
\label{eq:47}
\end{equation}
where $\Phi_{st}=M-M^*_{st}$ is the expection value of the $\sigma$-meson field at the saturation density $\rho_{st}$. 
The absolute values of these couplings are shown in Figs. \ref{fig:24} and \ref{fig:25}. 
It seems that there is an suppression of the odd couplings $b_l^*$. 
There may be the approximate discrete $\gamma_5$-symmetry at finite baryon density. 

Finally, we expand $\varepsilon (\rho ,\Phi ;\Lambda^\prime )$ around $M^*=M^*_{st}$ but put $\rho =0$, i.e., 
\begin{equation}
\varepsilon (\rho =0,\Phi ;\Lambda^\prime )=\sum_{l=0}{1\over{l!}}{{\partial^l }\over{\partial \Phi^l}}\varepsilon (\rho=0,\Phi ;\Lambda^\prime )\vert_{\Phi =\Phi_{st}}\Phi^l =\sum_{l=0}{{a_l^*}\over{g_s^l}} \Phi^l. 
\label{eq:48}
\end{equation}
The absolute values of these couplings are shown in Figs. \ref{fig:26} and \ref{fig:27}. 
We see that the coupling $a_l^*/g_s^l$ have similar features as in the case of $\rho =\rho_{st}$, except for $l=1$. 
Naturally, nonvanishing $a_1^*$ corresponds to the fact that $\Phi =\Phi_{st}$ is not the solution of the equation of motion (\ref{eq:Ad20}) at zero density. 

Although we did not show the results, the parameter set PS9 in Ref. 10) which has also small $M^*_{st}$ ($=0.5116M$) at $\rho_{st}$ yields the similar results as in the cases of PS10 and PS11. 
On the other hand, the parameter set PS4 in Ref. 10) which has large $M^*_{st}$ ($=0.8887M$) at $\rho_{st}$ yields no suppression of the odd coupling. 
However, the couplings $c_5$ and $c_6$ in PS4 are too large in comparison with the ones in PS10 and PS11, and it may not be the realistic parameter set. 

From these results, it seems that, there may be a approximate symmetry even at $\rho =0$ but nonvanishing $a_l^*$ causes the sift of the field $\phi$, yielding the enlargement of the other odd couplings and the large nucleon mass $M$. 
At the saturation density, the density contribution cancels the large $a_1^*$ and makes the nucleon mass smaller.

\section{Summary}

In this paper, we have studied the quantum hadrodynamics using the non-perturbative renormalization group equations. 
The results obtained here are summarized as follows. 

(1) At zero density, the NPRG equations are formulated in the quantum hadrodynamics ($\sigma$-nucleon model). 

(2) Approximate discrete $\gamma_5$-symmetry is studied. 
It is found that the approximate symmetry is retained to some extent through the whole evolutions in analogy of the exact one. 
However, the evolution makes the linear $\phi$-term large which causes the sift of the field $\phi$, yielding the enlargement of the other odd couplings and the large nucleon mass $M$. 

(3) Relations between the local-potential approximation in NPRG equations and the Hartree approximation are studied. 
The importance of the one-loop contribution is shown. 
The local-potential approximation contains not only the Hartree contribution but also the contributions of the Hartree-Fock, the RPA and the vertex corrections with vanishing external momentum. 

(4) The approximate discrete $\gamma_5$-symmetry is studied phenomenologically using the Hartree approximation. 
There may be the restoration of this symmetry at finite density, since the medium effects cancels the large linear $\phi$-term and makes the nucleon mass smaller. 

There are some interesting extensions of this work. 

(1) To formulate the NPRG equations at finite baryon density/finite temperature. 

(2) To include the $\omega$-meson to study the saturation mechanism. 

(3) To include the $\pi$-meson to study exact and approximate chiral symmetry. 

They are now under the studies. 

\section*{Acknowledgements}
We would like to thank K.-I.~Aoki, T.~Kohomura and H.~Yoneyama for useful discussions and suggestions. 
This work is supported in part by the Japanese Scientific Foundation (Grant No. 09740208). 





\bigskip

\bigskip

\bigskip

\hspace*{0cm}
\begin{tabular}{ccccccc}
    \hline 
    \      & $K$ & $M^*_{st}/M$ & $C_s^2$ & $C_v^2$ & $c_5M/g_s^5$ & $c_6M^2/g_s^6$ \\ 
    \hline
    \ PS10  & 300.0 & 0.5156 & 378.50 & 288.38 & 0.0024938 & -0.0018217 \\ 
    \ PS11  & 350.0 & 0.5442 & 358.11 & 270.07 & 0.0025532 & -0.0018579 \\ 
    \hline
\end{tabular}

\bigskip

\noindent
Table I; Parameter sets. $M^*_{st}$ is the effective nucleon mass at the saturation density. 
$K$ is the incompressibility of nuclear matter and shown in MeV. 
To get these parameter sets, we have used $\Lambda^\prime=$1.5GeV. 



\eject

\begin{figure}
\epsfxsize= 14cm 
\centerline{\epsfbox{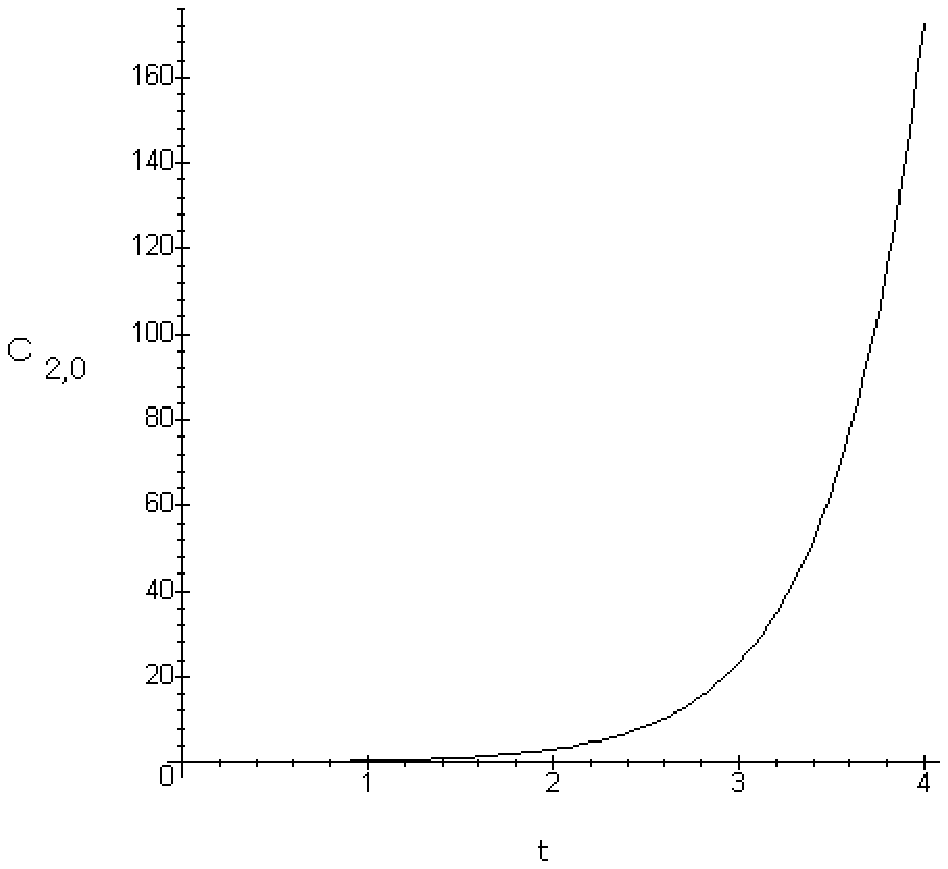}}
\caption{The evolution of the coupling $C_{2,0}$.}
\label{fig:1}
\end{figure}


\eject


\begin{figure}
\epsfxsize= 14cm 
\centerline{\epsfbox{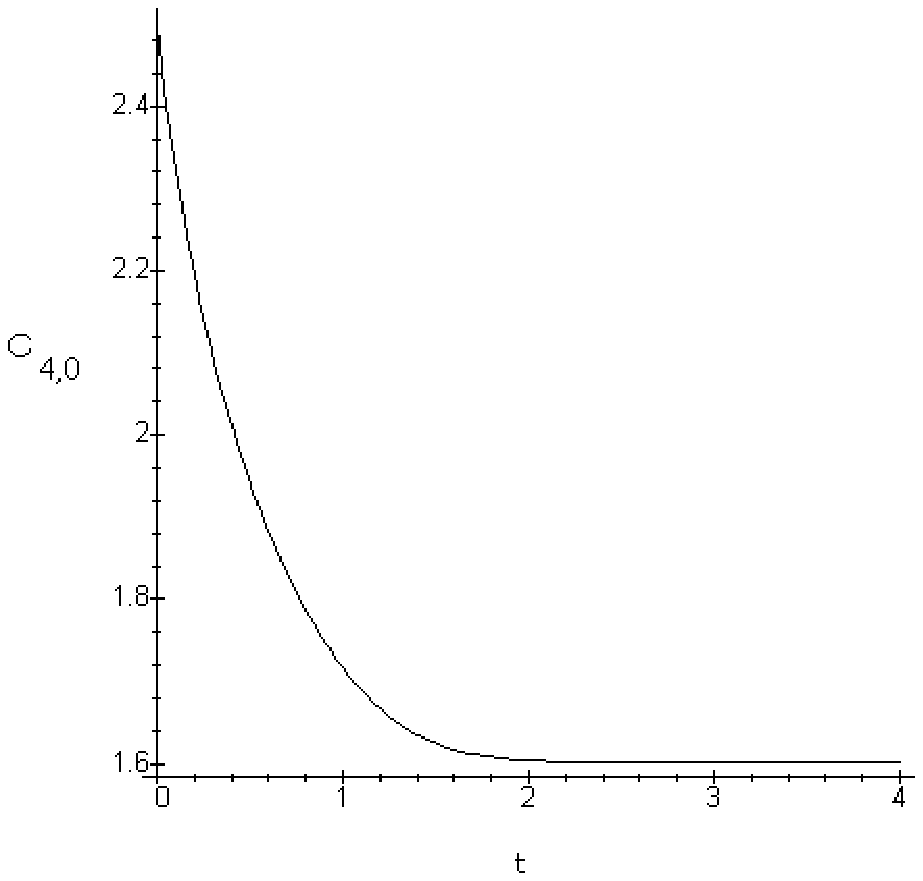}}
\caption{The evolution of the coupling $C_{4,0}$.}
\label{fig:2}
\end{figure}


\eject


\begin{figure}
\epsfxsize= 14cm 
\centerline{\epsfbox{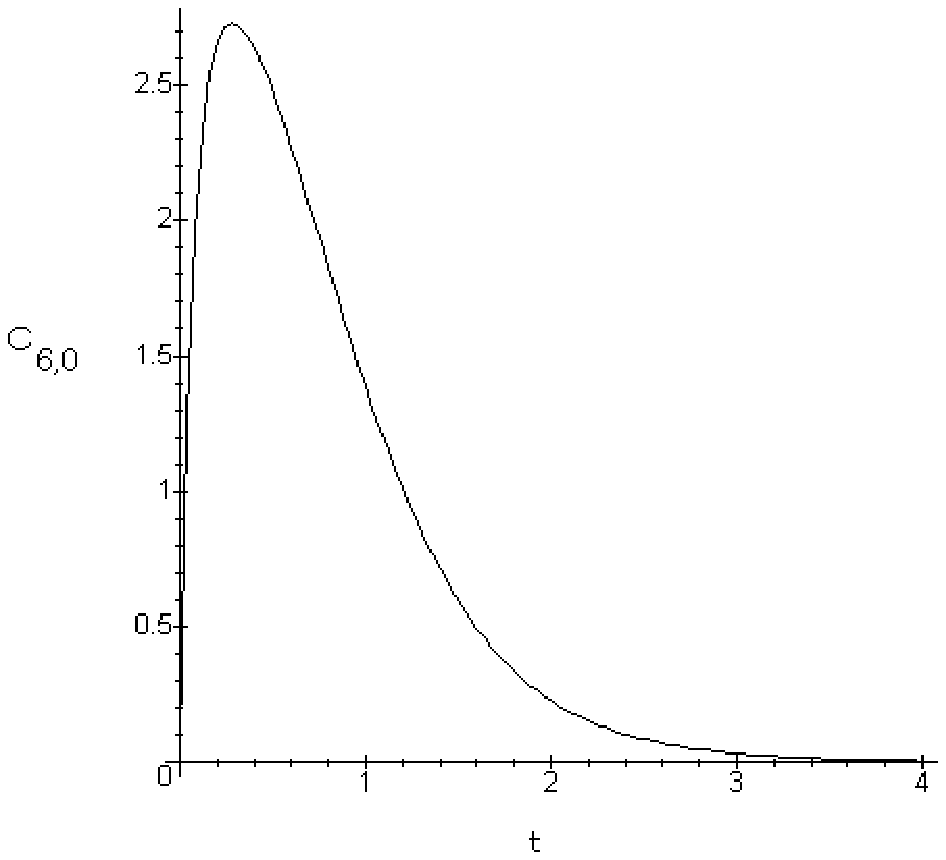}}
\caption{The evolution of the coupling $C_{6,0}$.}
\label{fig:3}
\end{figure}


\eject


\begin{figure}
\epsfxsize= 14cm 
\centerline{\epsfbox{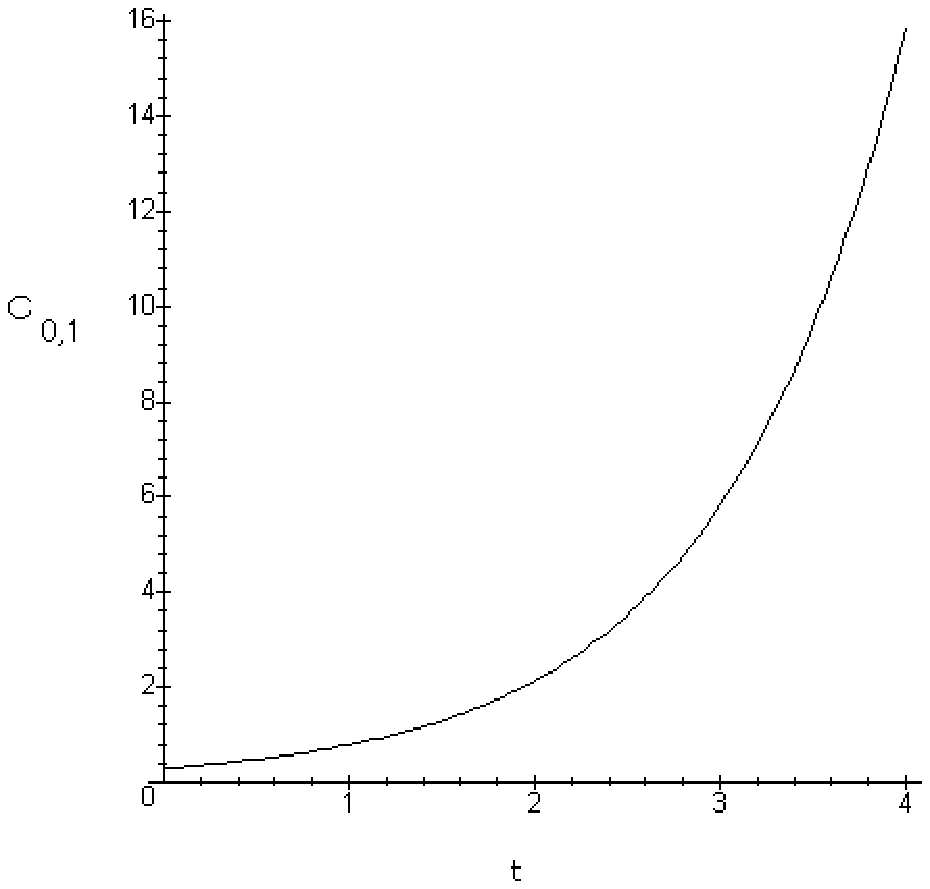}}
\caption{The evolution of the coupling $C_{0,1}$.}
\label{fig:4}
\end{figure}


\eject


\begin{figure}
\epsfxsize=14cm 
\centerline{\epsfbox{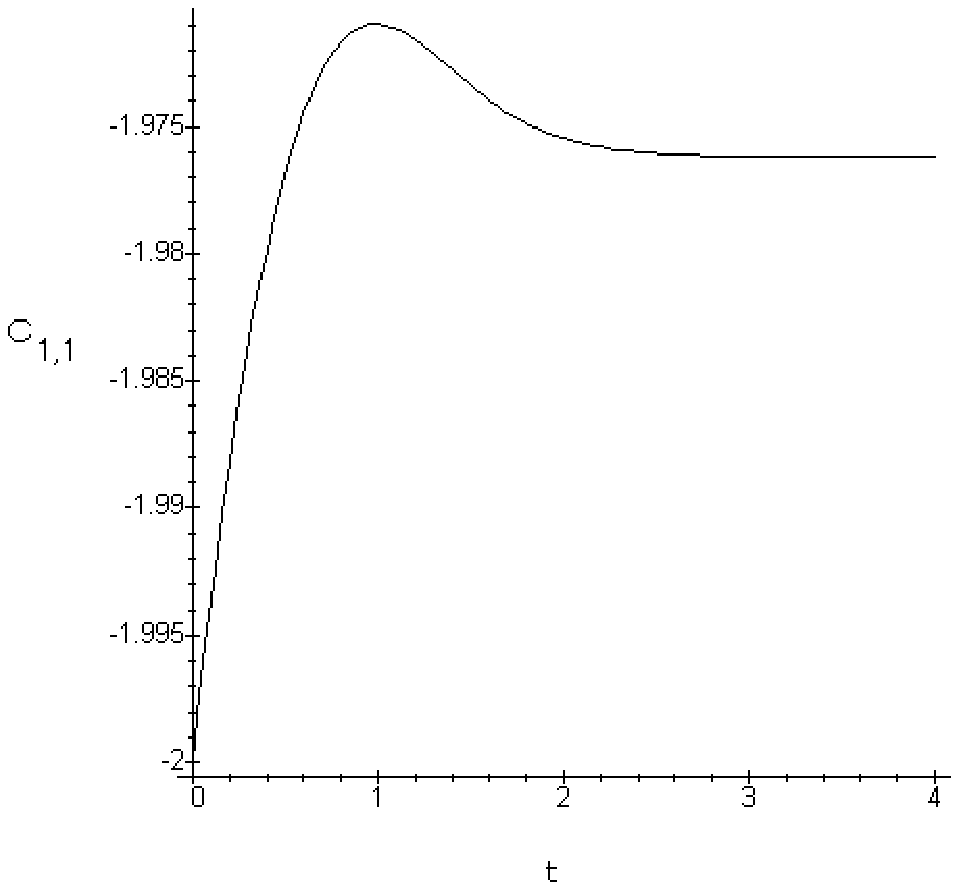}}
\caption{The evolution of the coupling $C_{1,1}$.}
\label{fig:5}
\end{figure}


\eject


\begin{figure}
\epsfxsize= 14cm 
\centerline{\epsfbox{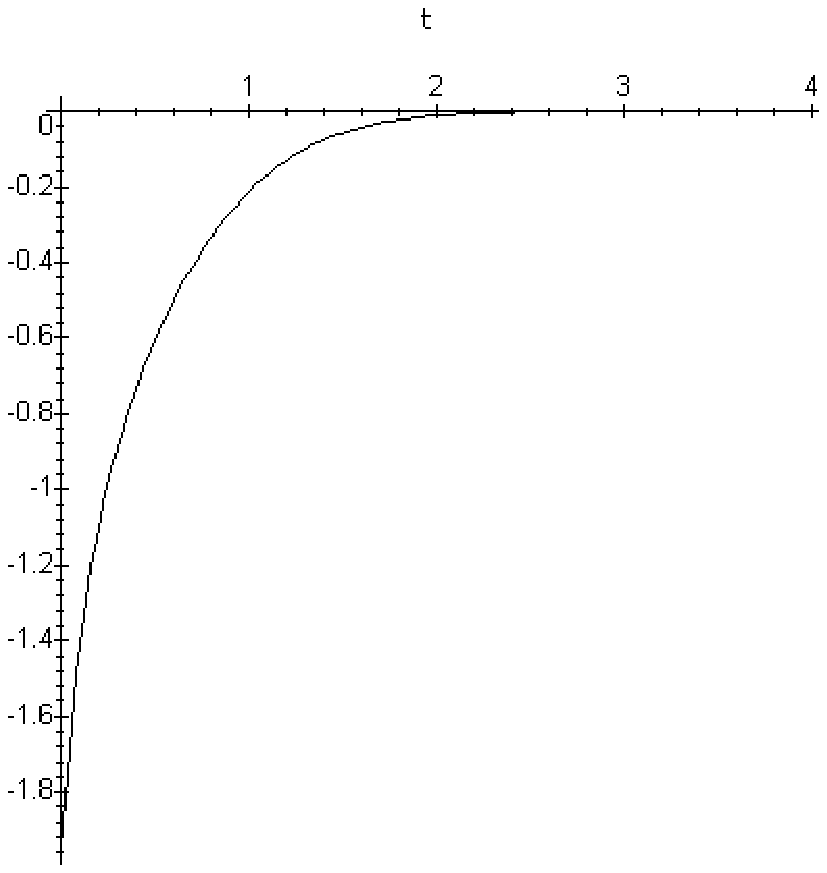}}
\caption{Bosonic contribution for the evolution of the coupling $C_{4,0}$}
\label{fig:6}
\end{figure}


\eject


\begin{figure}
\epsfxsize= 14cm 
\centerline{\epsfbox{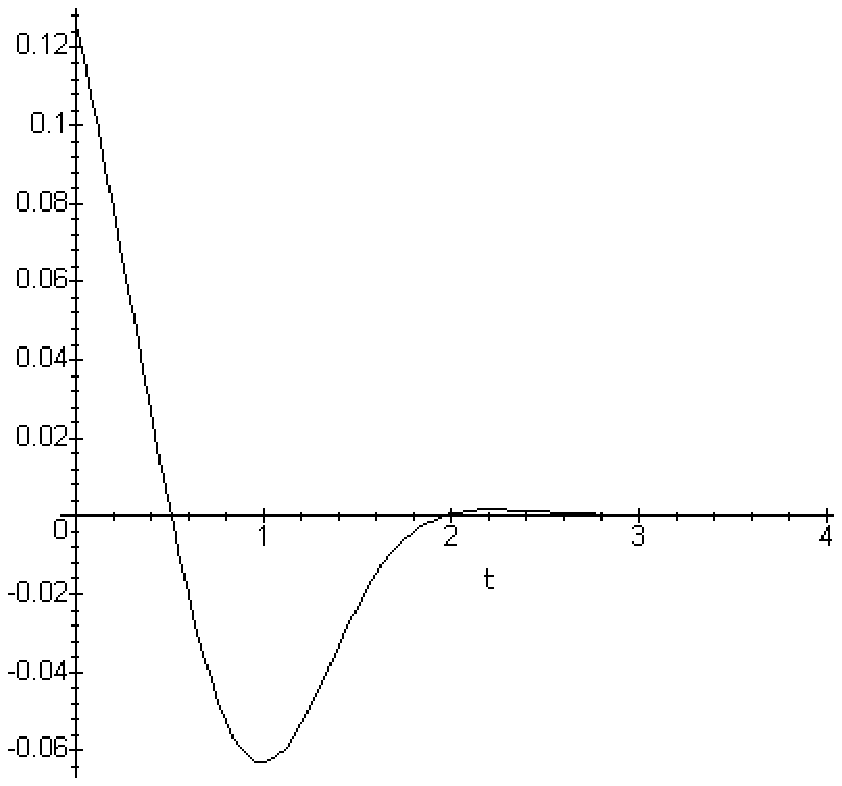}}
\caption{Fermionic contribution for the evolution of the coupling $C_{4,0}$}
\label{fig:7}
\end{figure}

\eject

\begin{figure}
\epsfxsize= 14cm 
\centerline{\epsfbox{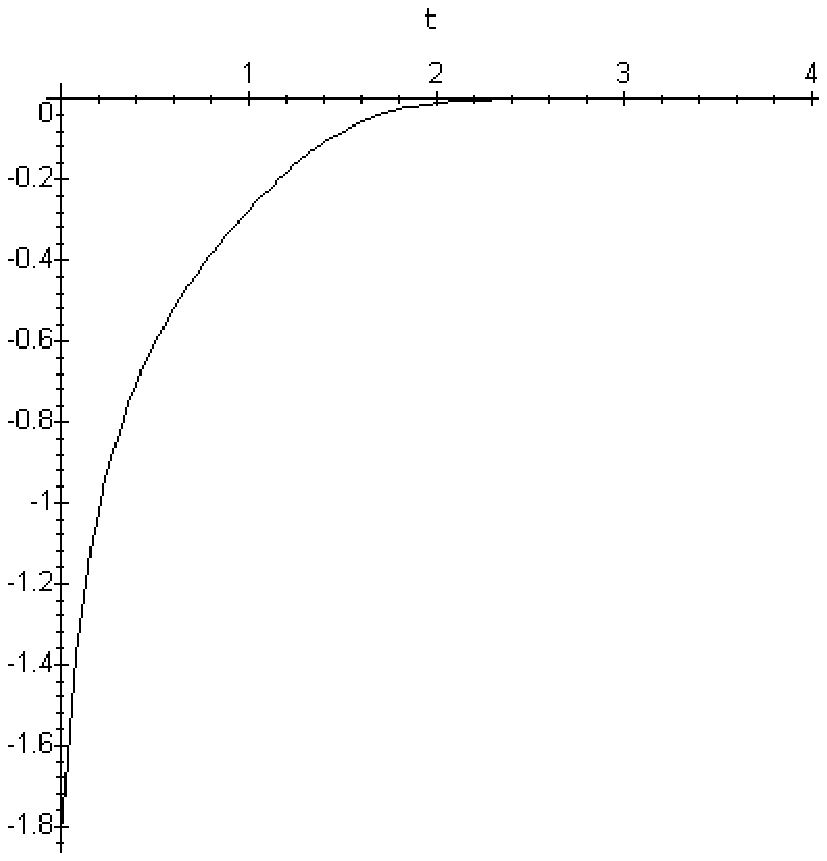}}
\caption{Total interaction contribution for the evolution of the coupling $C_{4,0}$}
\label{fig:8}
\end{figure}

\eject

\begin{figure}
\epsfxsize= 14cm 
\centerline{\epsfbox{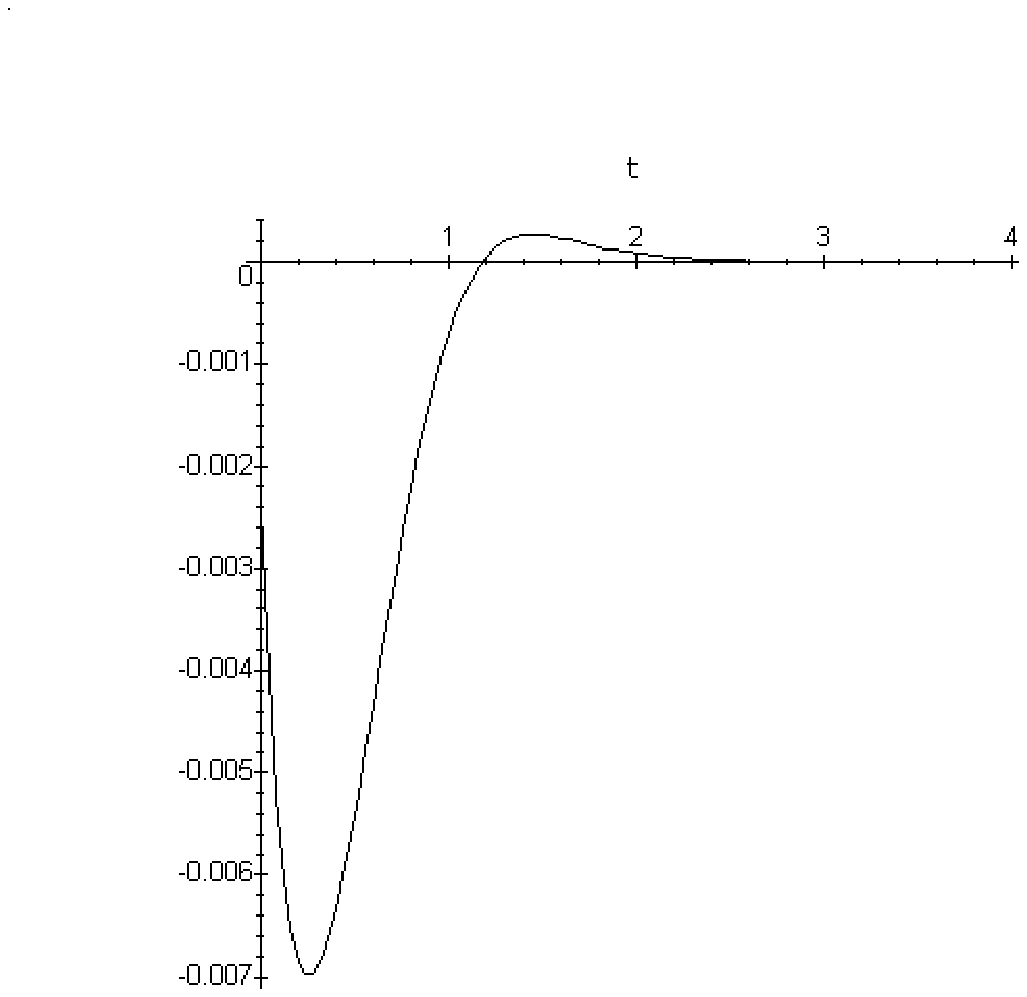}}
\caption{Bosonic contribution for the evolution of the coupling $C_{1,1}$}
\label{fig:9}
\end{figure}

\eject

\begin{figure}
\epsfxsize= 14cm 
\centerline{\epsfbox{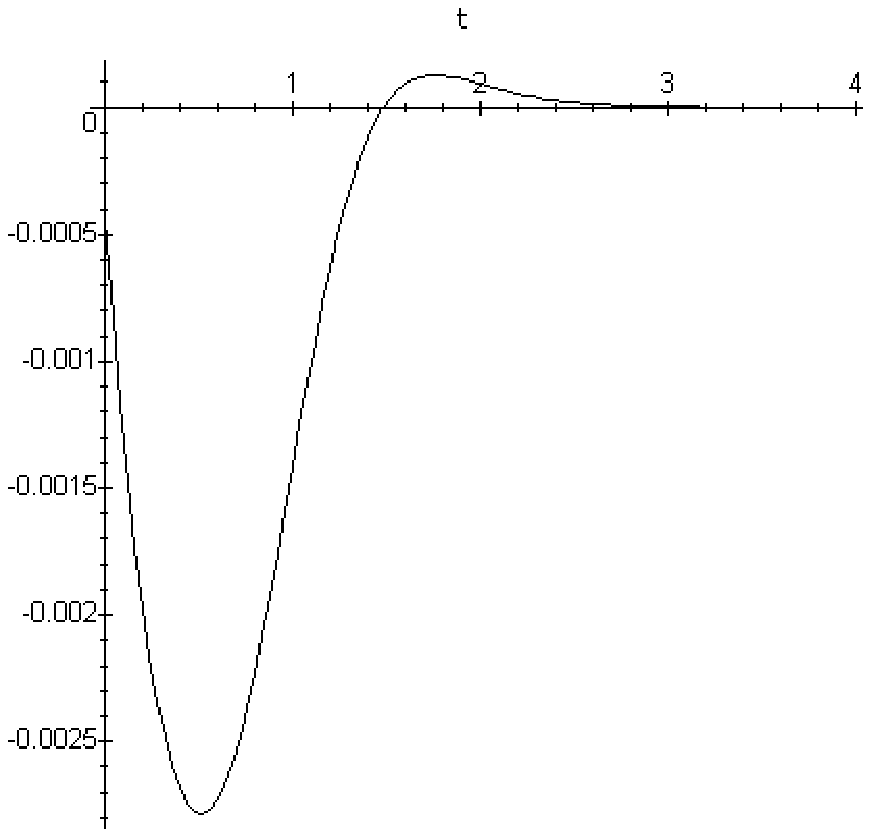}}
\caption{Fermionic contribution for the evolution of the coupling $C_{1,1}$}
\label{fig:10}
\end{figure}

\eject

\begin{figure}
\epsfxsize= 14cm 
\centerline{\epsfbox{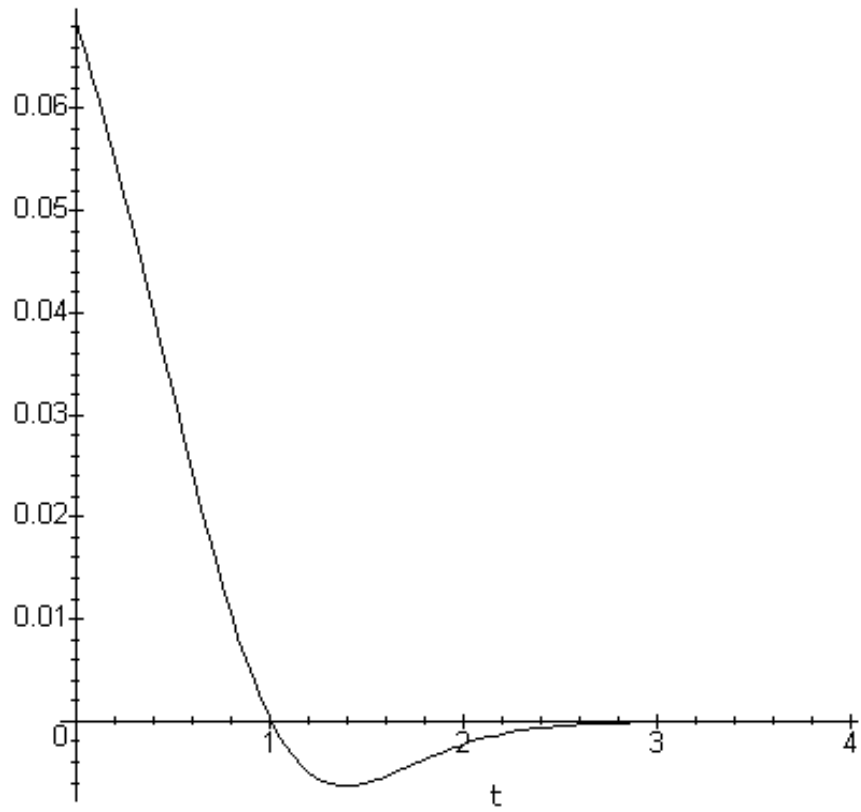}}
\caption{BF contribution for the evolution of the coupling $C_{1,1}$}
\label{fig:11}
\end{figure}

\eject

\begin{figure}
\epsfxsize= 14cm 
\centerline{\epsfbox{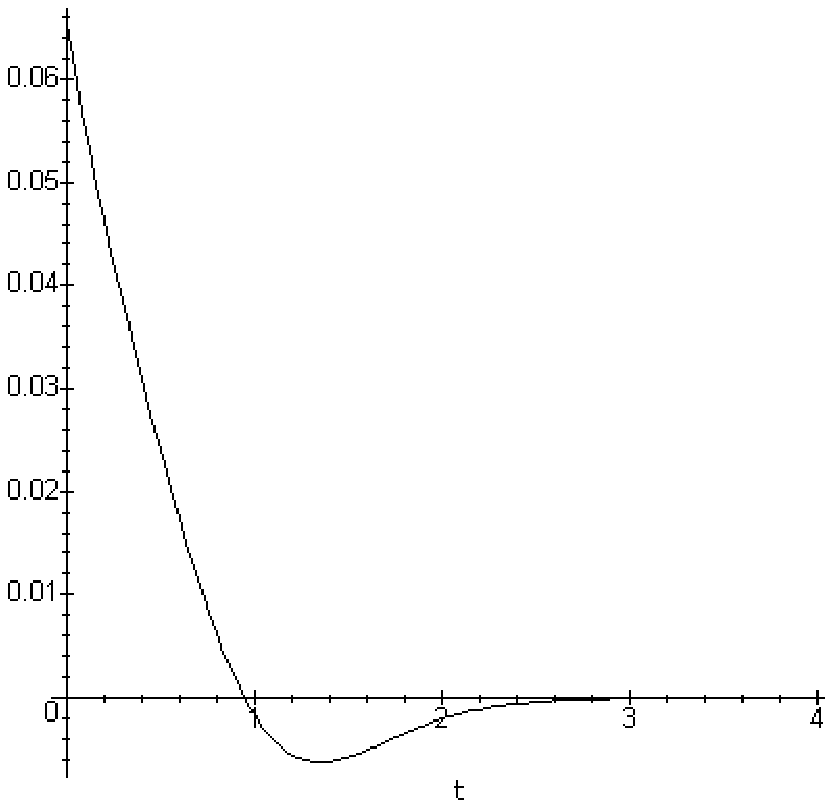}}
\caption{Total interaction contribution for the evolution of the coupling $C_{1,1}$}
\label{fig:12}
\end{figure}

\eject

\begin{figure}
\epsfxsize= 12cm 
\centerline{\epsfbox{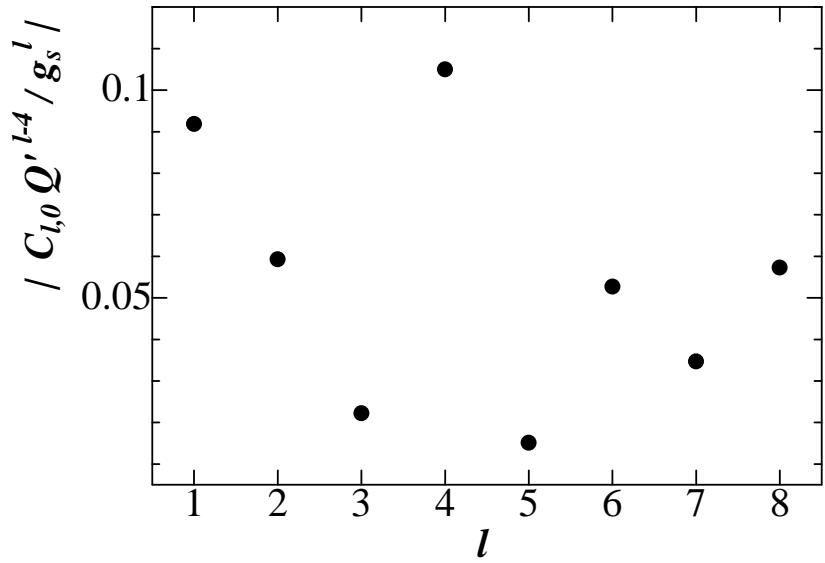}}
\caption{The absolute value of $C_{l,0}(Q^\prime )^{l-4}/g_s^l$ at $t=4$. }
\label{fig:13}
\end{figure}

\eject

\begin{figure}
\epsfxsize= 12cm 
\centerline{\epsfbox{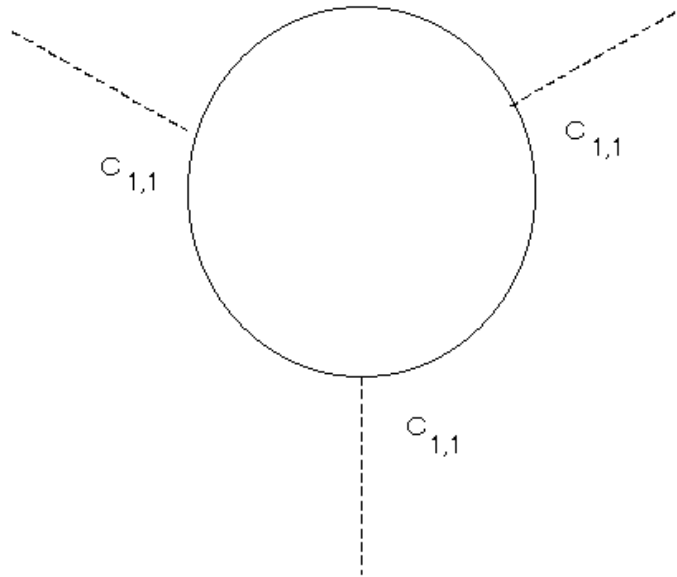}}
\caption{The one-loop contribution with the Yukawa coupling $C_{1,1}$ for $C_{3,0}$.
The solid and dotted lines are fermion and boson lines, respectively. }
\label{fig:14}
\end{figure}

\eject

\begin{figure}
\epsfxsize= 14cm 
\centerline{\epsfbox{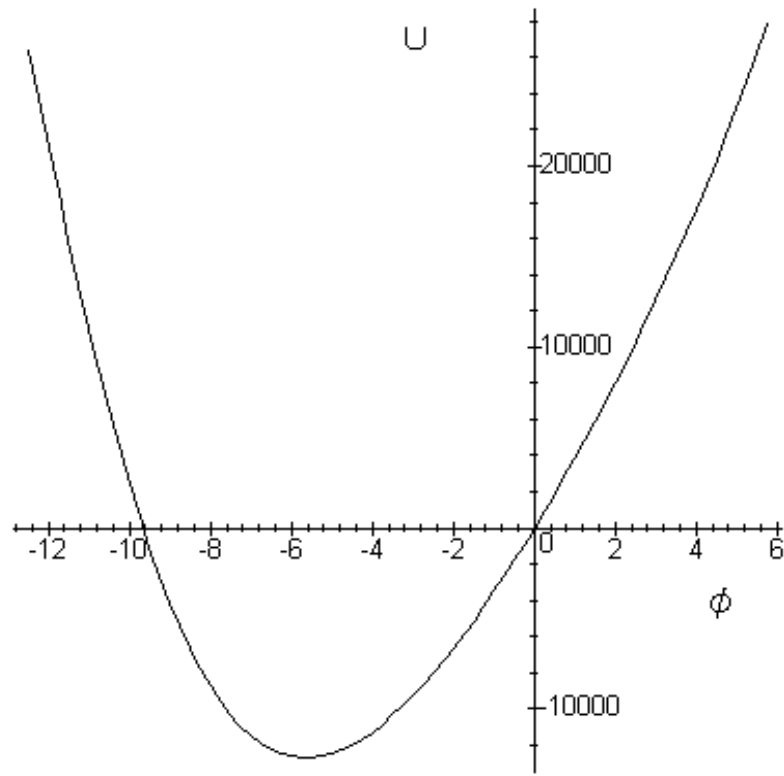}}
\caption{Effective generalized potential at $\sigma =0$ and $t=4$. }
\label{fig:15}
\end{figure}

\eject

\begin{figure}
\epsfxsize= 14cm 
\centerline{\epsfbox{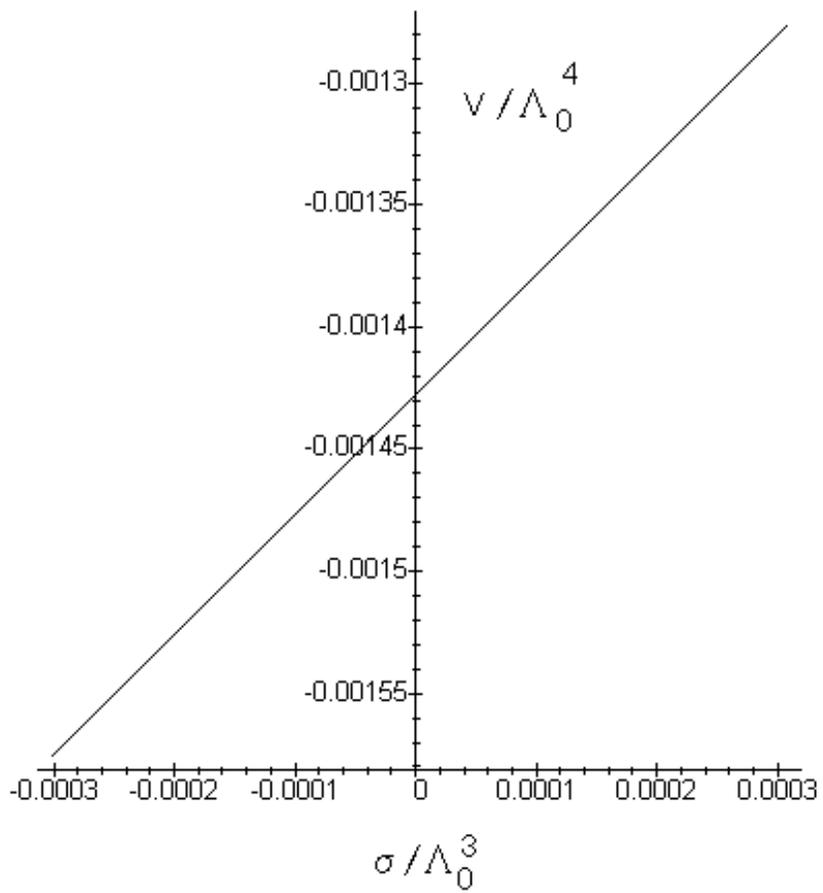}}
\caption{$V$ at $\phi =0.1\Lambda_0$ and $t=4$.}
\label{fig:16}
\end{figure}

\eject

\begin{figure}
\epsfxsize= 14cm 
\centerline{\epsfbox{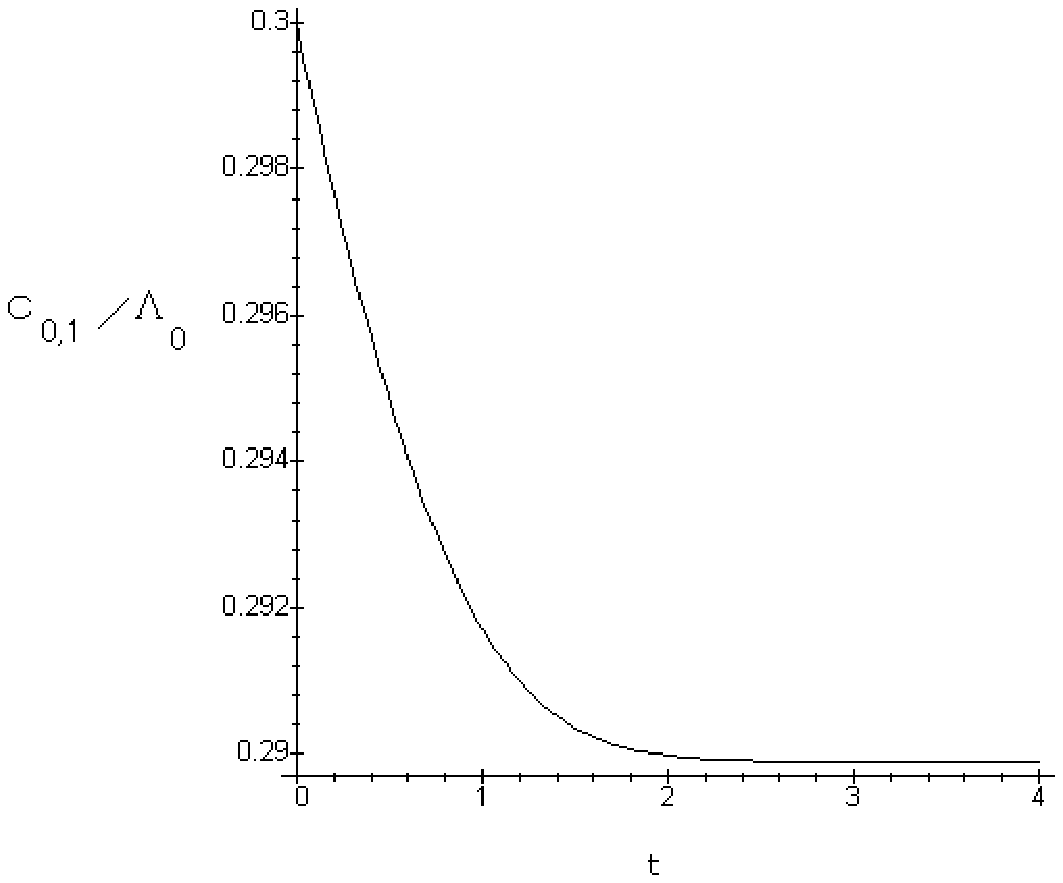}}
\caption{The evolution of the coupling $C_{0,1}$ without the caret.}
\label{fig:17}
\end{figure}

\clearpage

\begin{figure}
\epsfxsize= 14cm 
\centerline{\epsfbox{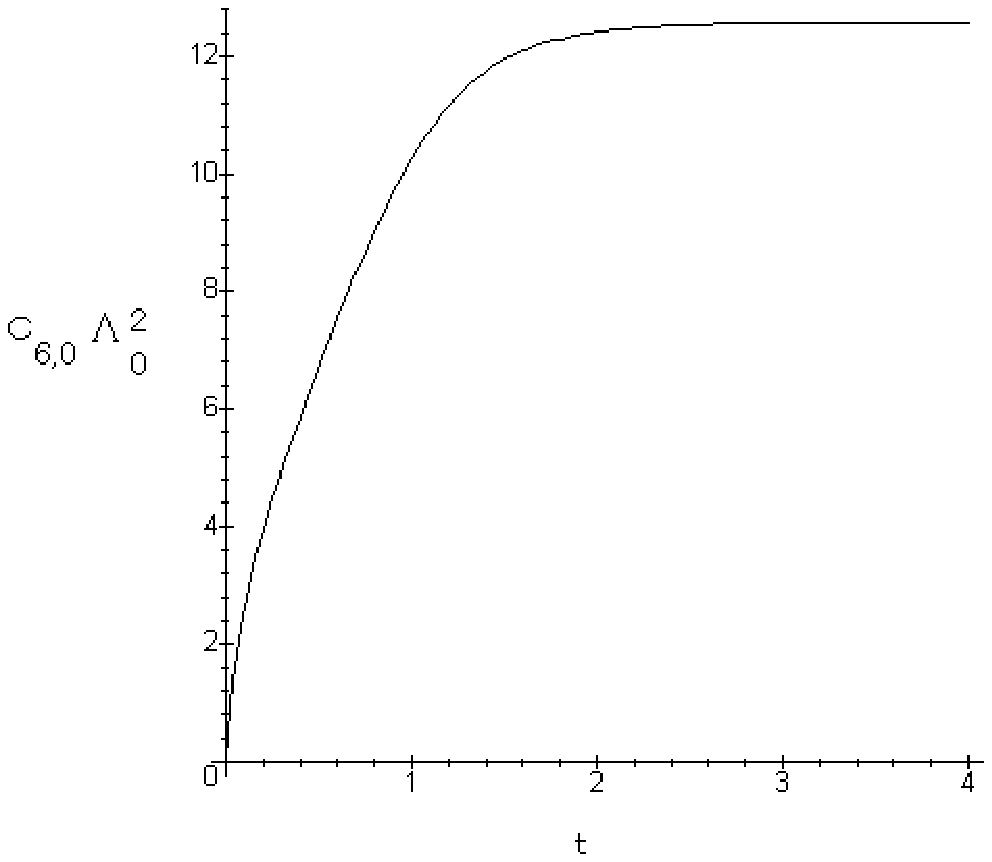}}
\caption{The evolution of the coupling $C_{6,0}$ without the caret. }
\label{fig:18}
\end{figure}
\eject

\begin{figure}
\epsfxsize= 14cm 
\centerline{\epsfbox{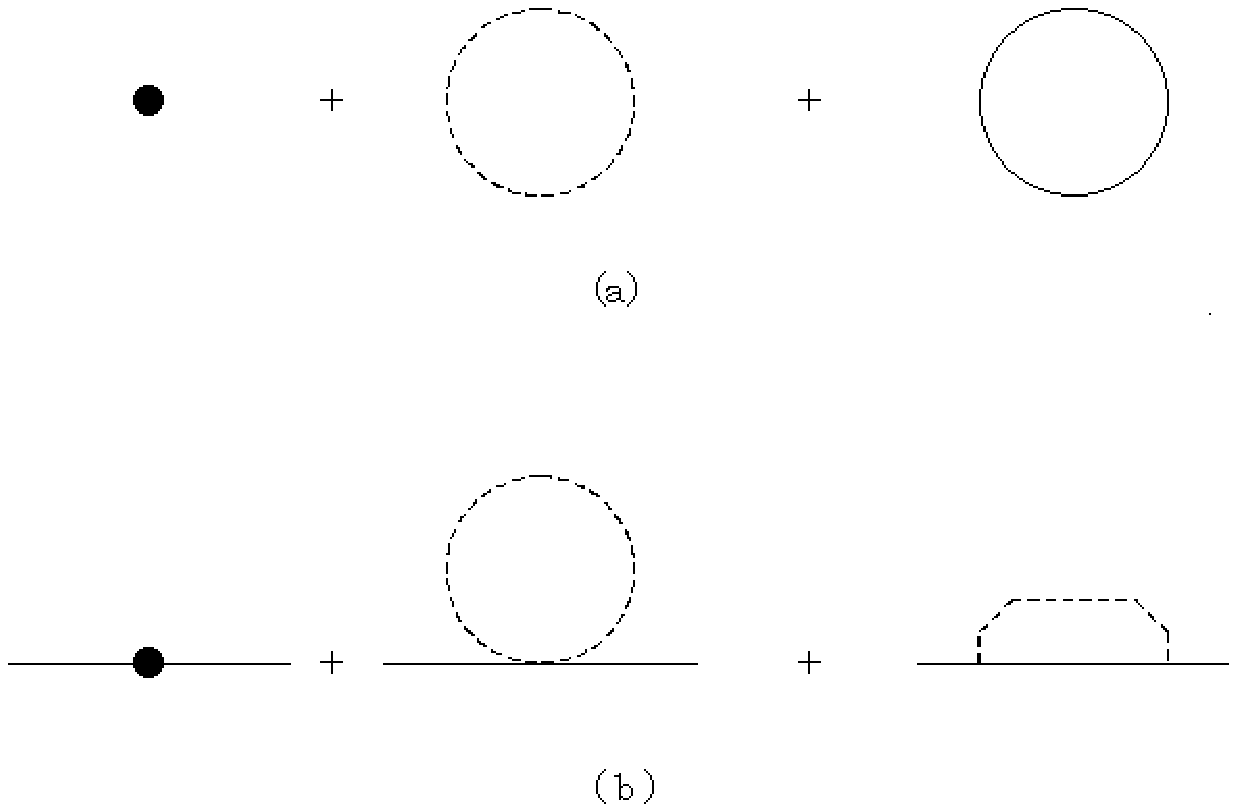}}
\caption{The graphical expression for $V^{(0)}$ ((a))and $V^{(1)}$ ((b)). }
\label{fig:19}
\end{figure}

\eject

\begin{figure}
\epsfxsize= 12cm 
\centerline{\epsfbox{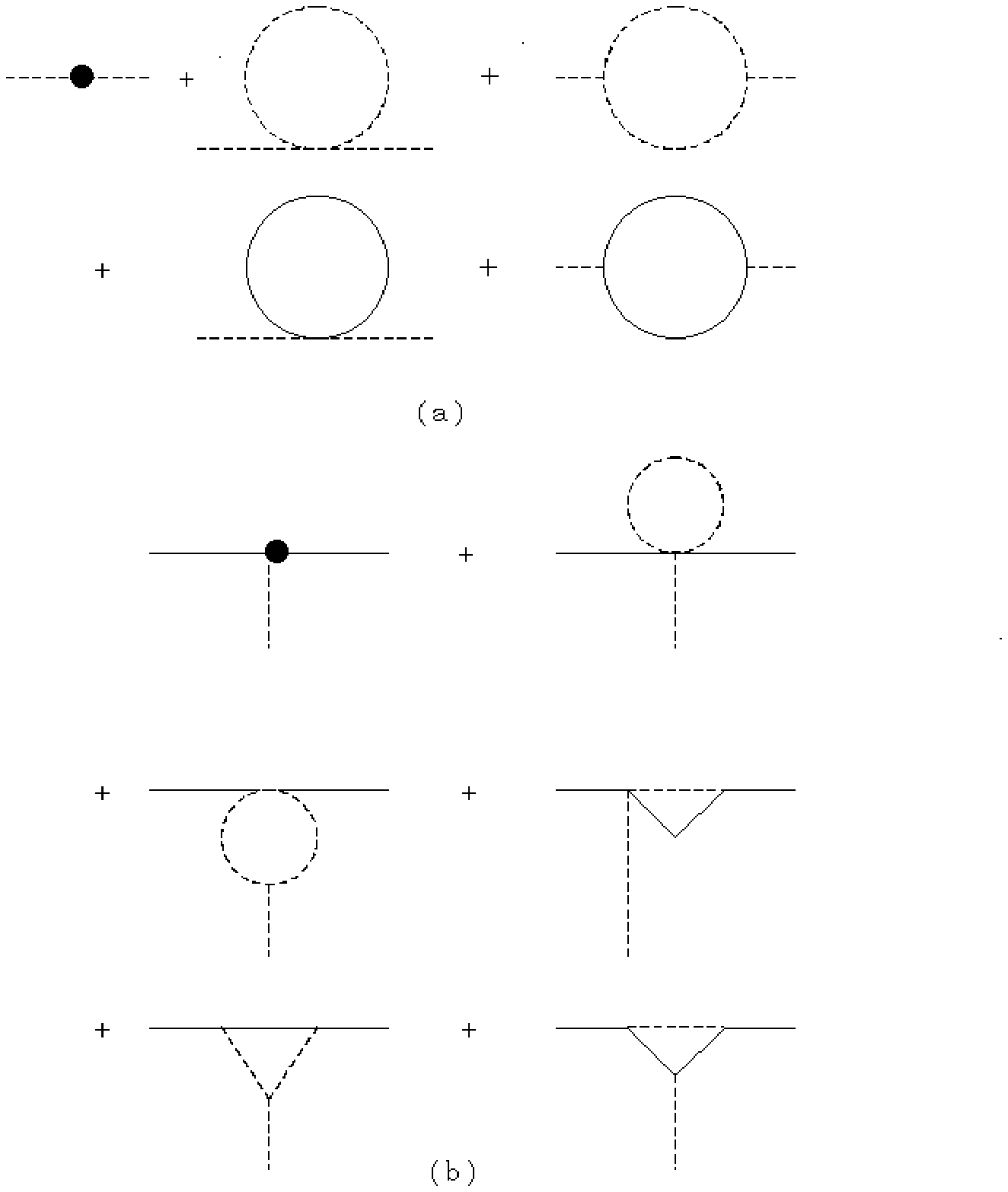}}
\caption{The graphical expression for $V^{(0)}_{\phi\phi}$ ((a)) and $V^{(1)}_{\phi}$ ((b)).}
\label{fig:20}
\end{figure}

\eject

\begin{figure}
\epsfxsize= 14cm 
\centerline{\epsfbox{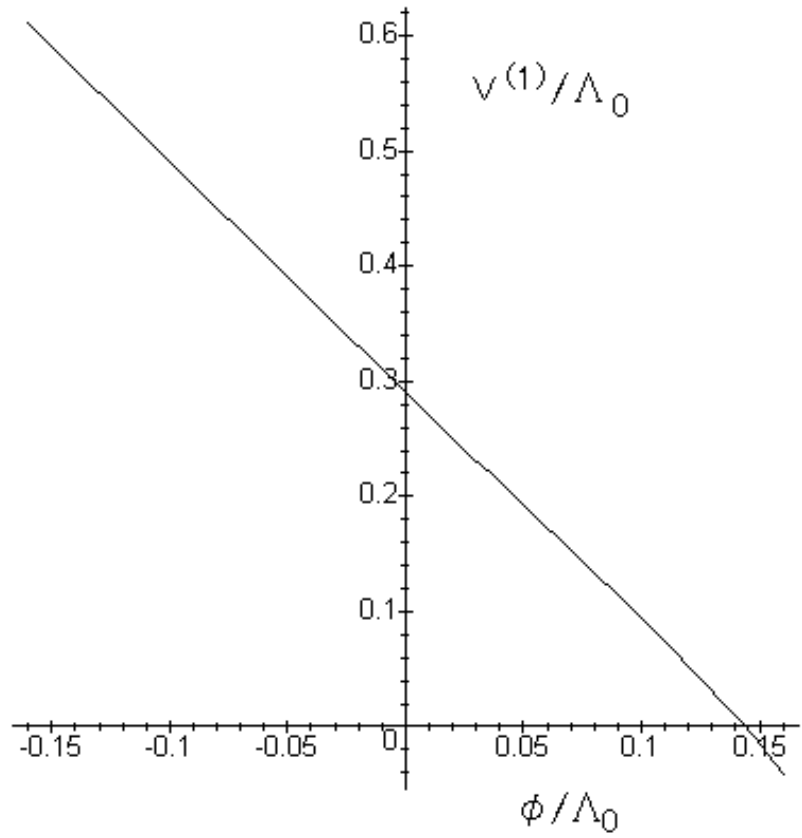}}
\caption{$V^{(1)}$ at $t=4$. }
\label{fig:21}
\end{figure}

\eject

\begin{figure}
\epsfxsize= 14cm 
\centerline{\epsfbox{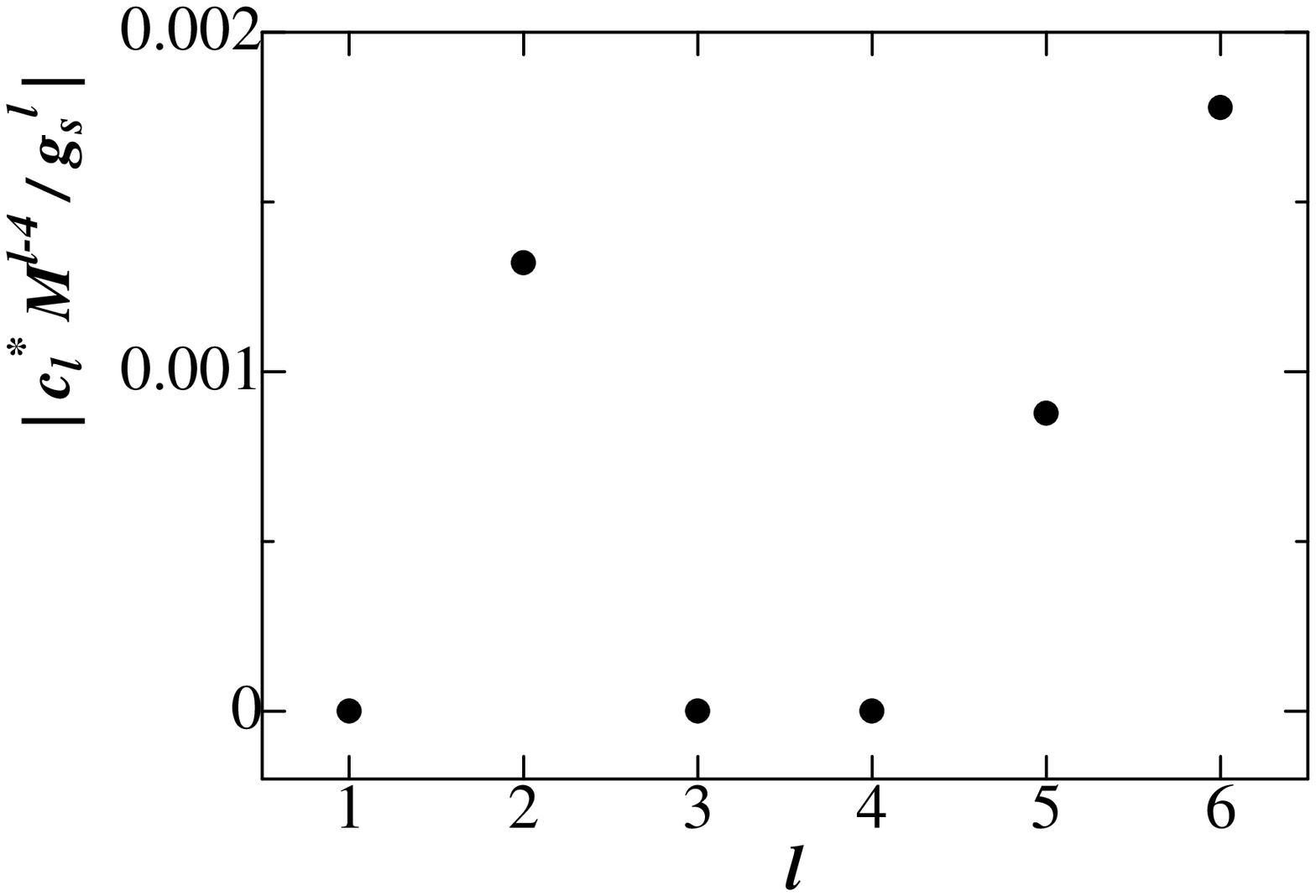}}
\caption{The absolute value of $c^*_{l}M^{l-4}/g_s^l$ in PS10. }
\label{fig:22}
\end{figure}

\eject

\begin{figure}
\epsfxsize= 14cm 
\centerline{\epsfbox{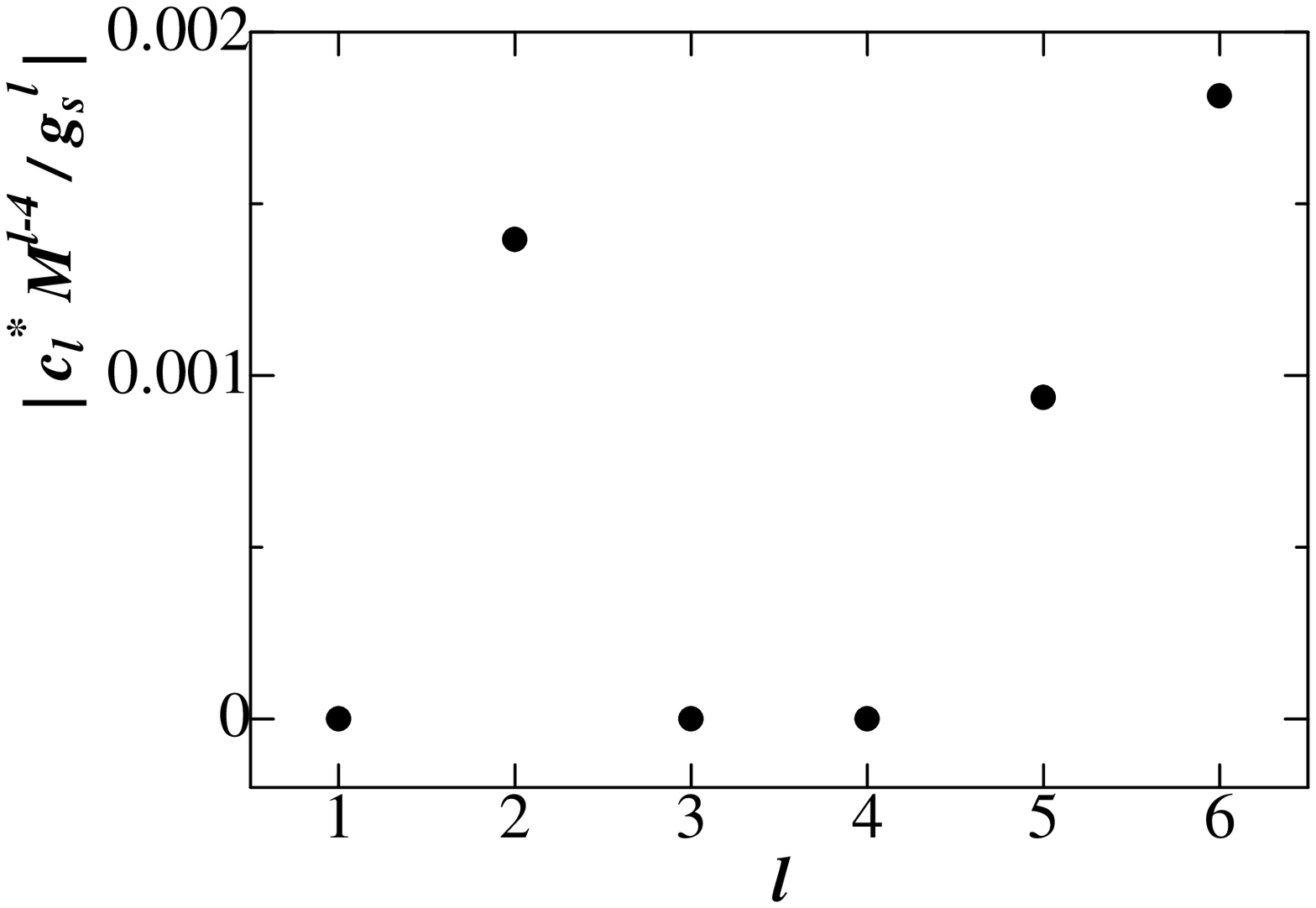}}
\caption{The absolute value of $c^*_{l}M^{l-4}/g_s^l$ in PS11. }
\label{fig:23}
\end{figure}

\eject

\begin{figure}
\epsfxsize= 14cm 
\centerline{\epsfbox{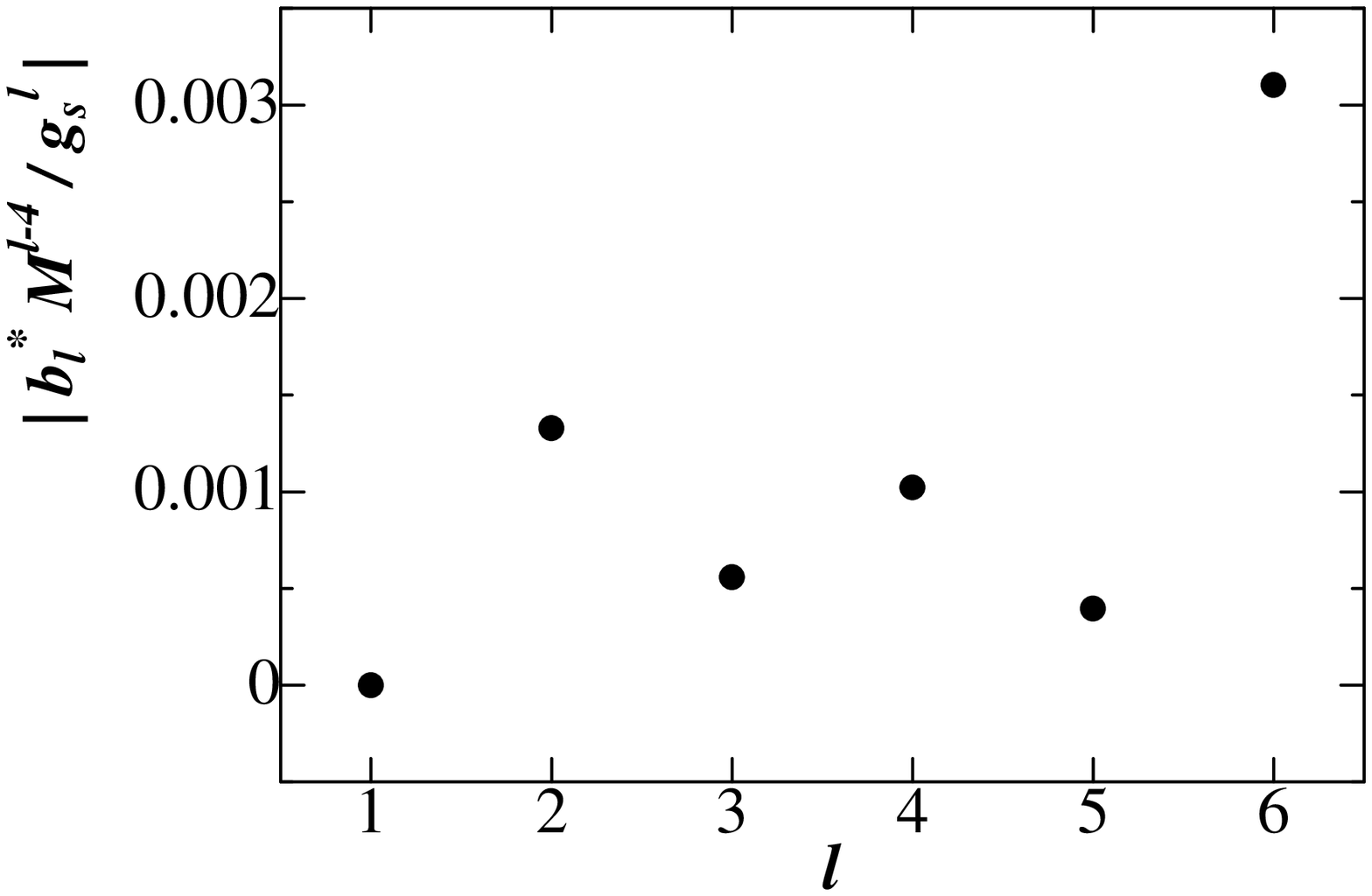}}
\caption{The absolute value of $b^*_{l}M^{l-4}/g_s^l$ in PS10. }
\label{fig:24}
\end{figure}

\eject

\begin{figure}
\epsfxsize= 14cm 
\centerline{\epsfbox{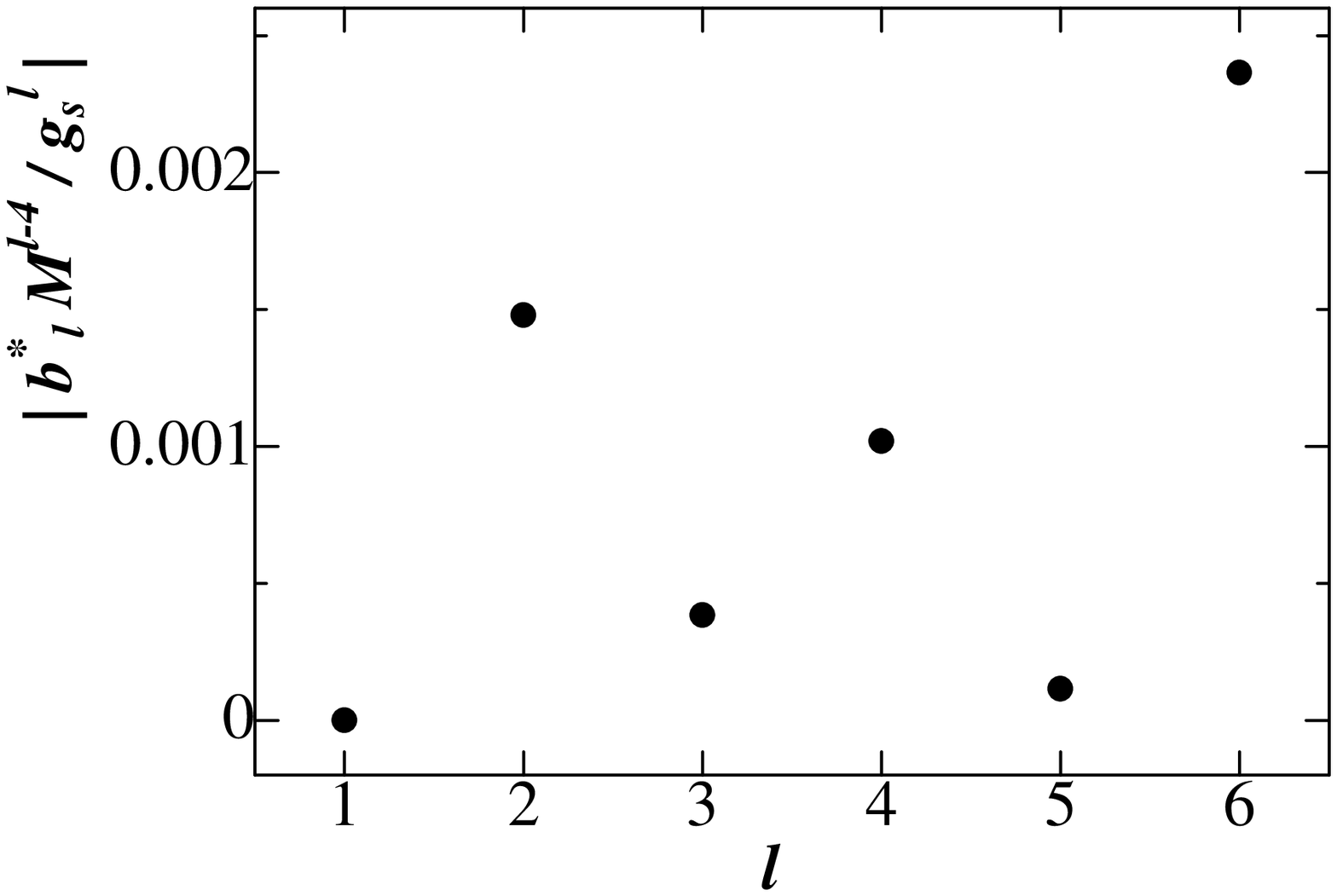}}
\caption{The absolute value of $b^*_{l}M^{l-4}/g_s^l$ in PS11. }
\label{fig:25}
\end{figure}

\eject

\begin{figure}
\epsfxsize= 14cm 
\centerline{\epsfbox{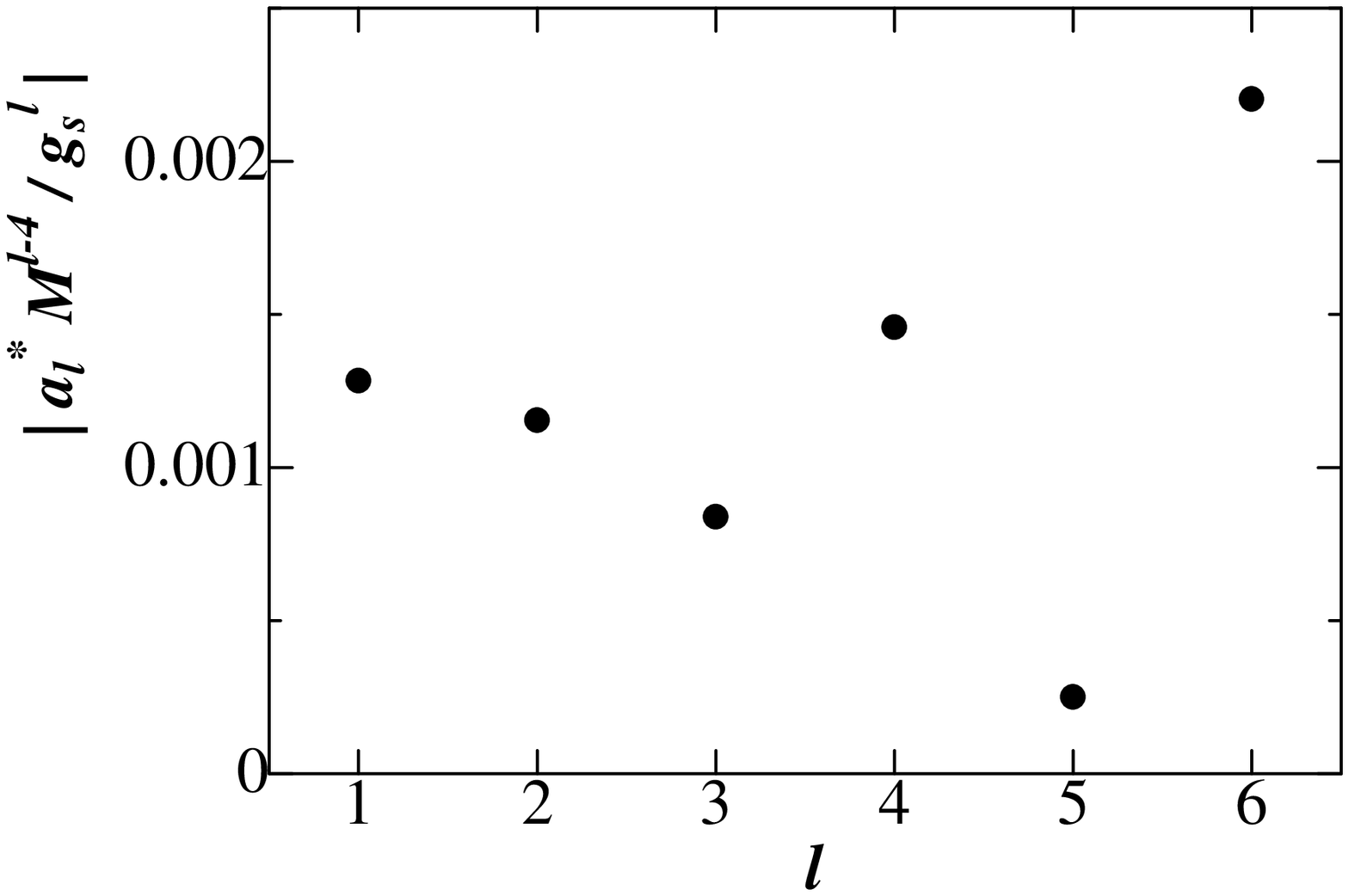}}
\caption{The absolute value of $a^*_{l}M^{l-4}/g_s^l$ in PS10. }
\label{fig:26}
\end{figure}

\eject

\begin{figure}
\epsfxsize= 14cm 
\centerline{\epsfbox{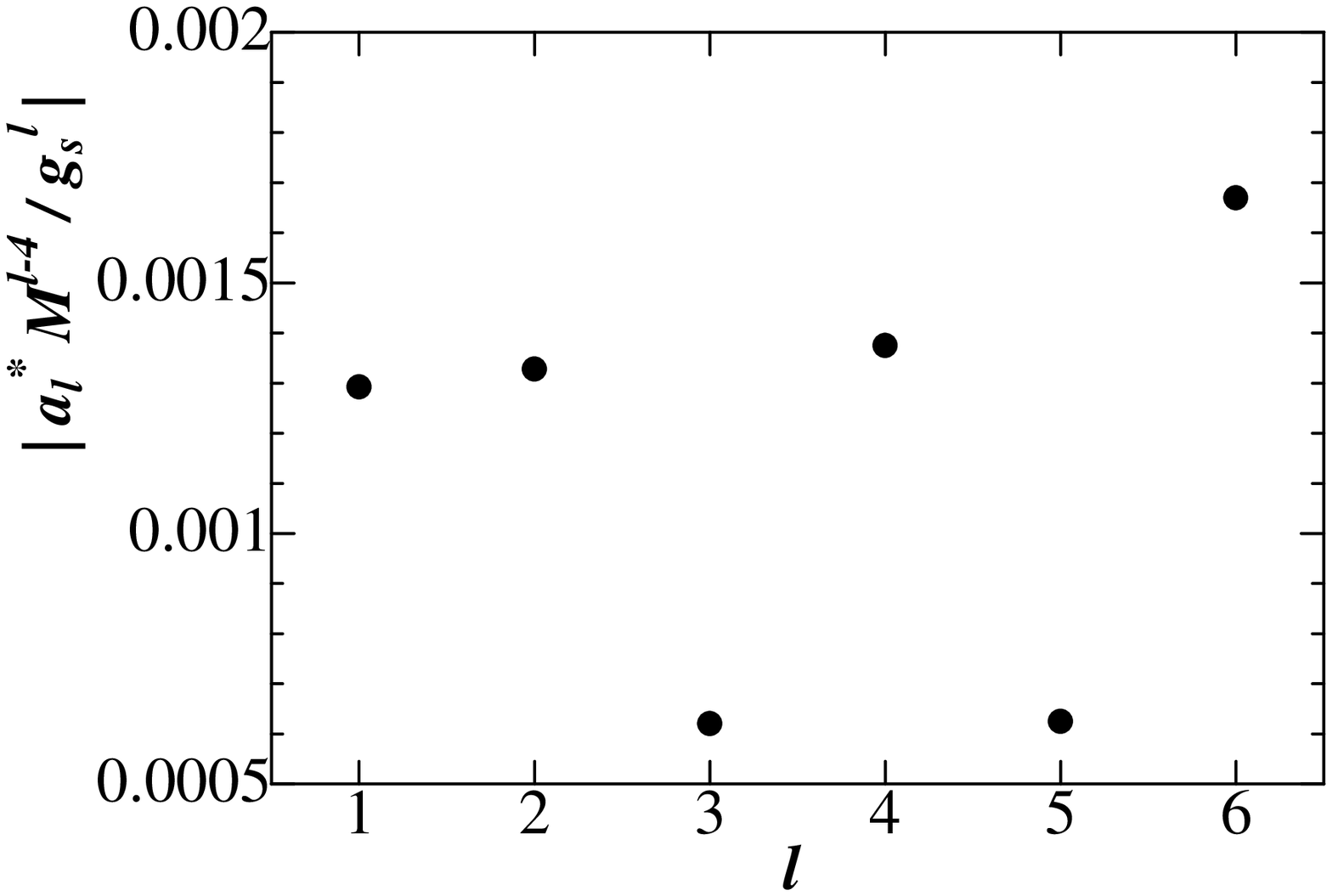}}
\caption{The absolute value of $a^*_{l}M^{l-4}/g_s^l$ in PS11. }
\label{fig:27}
\end{figure}



\begin{thebibliography}{99}
\bibitem{rf:Walecka}
J.~D.~Walecka, Ann. of Phys. {\bf 83} (1974), 491.  
\bibitem{rf:Serot}
B.~D.~Serot and J.~D.~Walecka, "The Relativistic Nuclear Many-Body 

Problem" in {\it Advances in nuclear physics}, vol. {\bf 16}, ed. by J~.W.~Negele 

and E.~Vogt (Plenum Press, New York, 1986).
\bibitem{rf:Chin}
S.~A.~Chin, Phys. Lett. {\bf 62B} (1976), 263; 
S.~A.~Chin, Ann. of Phys. {\bf 108} (1977), 301. 
\bibitem{rf:Kohmura}
T.~Kohmura, Y.~Miyama, T.~Nagai, S.~Ohnaka, J.~Da~Provid{\^{e}}ncia, and T.~Kodama, Phys. Lett. {\bf B226}~(1989), 207. 
\bibitem{rf:Furn}
R.~J.~Furnstahl and C.~J.~Horowitz, Nucl. Phys. {\bf A485}~(1988), 632.
\bibitem{rf:Kurasawa}
H.~Kurasawa and T.~Suzuki, Nucl. Phys. {\bf A445}~(1985), 685; 

\noindent
Nucl. Phys. {\bf A490}~(1988), 571. 
\bibitem{rf:Cohen}
T.~D.~Cohen, Phys. Lett. {\bf B211}~(1988), 384. 
\bibitem{rf:Lepage}
G.~P.~Lepage, in "From actions to answers", {\it Proceedings of the 1989 

theoretical advanced study institute in elementary particle physics}, ed. by 

T.~DeGrand and D.~Toussaint (World Scientific, Singapore, 1990), p.483. 
\bibitem{rf:Kouno1}
H.~Kouno, T.~Mitsumori, Y.~Iwasaki, K.~Sakamoto, N.~Noda, K.~Koide, 

A.~Hasegawa and M.~Nakano, Prog. Theor. Phys. {\bf 97} (1997), 91. 
\bibitem{rf:Kouno2}
H.~Kouno, K.~Sakamoto, Y.~Iwasaki, N.~Noda, T~. Mitsumori, K.~Koide, 

A.~Hasegawa and M.~Nakano, Prog. Theor. Phys. {\bf 98} (1997), 1123. 
\bibitem{rf:Kouno3}
H.~Kouno, K.~Koide, N.~Noda, K.~Sakamoto, Y.~Iwasaki, T.~Mitsumori, 

A.~Hasegawa and M.~Nakano, Prog. Theor. Phys. {\bf 99} (1998), 395. 
\bibitem{rf:Wilson}
K.~G.~Wilson, Rev. Mod. Phys. {\bf 47}~(1975), 773;  
K.~G.~Wilson and J.~Kogut, Phys. Rep. {\bf 12}~(1974), 75. 
\bibitem{rf:Wegner}
F.~J.~Wegner and A. Houghton, Phys. Rev. {\bf A8}~(1973), 401.
\bibitem{rf:Nicoll}
J.~F.~Nicoll, T.~S.~Chang and H.~E.~Stanley, Phys. Rev. Lett. {\bf 33}~(1974), 540.
\bibitem{rf:Polch}
J.~Polchinski, Nucl. Phys. {\bf B231}~(1984), 269.
\bibitem{rf:Hasen}
A.~Hasenfraz and P.~Hasenfraz, Nucl. Phys. {\bf B270}~(1986), 687. 
\bibitem{rf:Wetterich}
C.~Wetterich, Phys. Lett. {\bf B301}~(1993), 90.
\bibitem{rf:Clark}
T.E.~Clark, B.~Haeri and S.T.~Love, 
Nucl. Phys. {\bf B402} (1993), 628.   
\bibitem{rf:Morris}
T.~R.~Morris, J.~Mod.~Phys. {\bf A9}~(1994), 2411. 
\bibitem{rf:Aoki1}
K-I.~Aoki, K.~Morikawa, W.~Souma, J-I.~Sumi and H.~Terao, 
Prog. Theor. Phys., {\bf 95}~(1996), 409. 
\bibitem{rf:Aoki2}
K-I.~Aoki, K.~Morikawa, J-I.~Sumi, H.~Terao and M. Tomoyose, 
Prog. Theor. Phys., {\bf 97}~(1997), 479. 
\bibitem{rf:Aoki3}
K-I.~Aoki, K.~Morikawa, W.~Souma, J-I.~Sumi and H.~Terao, 
Prog. Theor. Phys., {\bf 99}~(1998), 451. 
\bibitem{rf:Furn2}
R.~J.~Furnstahl, R.~J.~Perry and B.~D.~Serot, Phys. Rev. {\bf C40}~(1989), 321.
\bibitem{rf:Nakano1}
M.~Nakano and A. Hasegawa, Phys. Rev. C{\bf 43}~(1991), 618. 
\bibitem{rf:Nakano2}
M.~Nakano, A. Hasegawa, H. Kouno and K. Koide, Phys. Rev. C{\bf 49}~(1994), 3061. 
\bibitem{rf:Nakano3}
M.~Nakano, T. Mitsumori, M. Muraki, K. Koide, H. Kouno and A. Hasegawa, Phys. Rev. C{\bf 49}~(1994), 3076. 
\bibitem{rf:Hasegawa1}
A.~Hasegawa, K.~Koide, T.~Mitsumori, M.~Muraki,  H.~Kouno and M.~Nakano, 
Prog. Theor. Phys. {\bf 92} (1994), 331. 
\bibitem{rf:Hasegawa2}
A.~Hasegawa, T.~Mitsumori, M.~Muraki, K.~Koide, H.~Kouno and M.~Nakano, 
Prog. Theor. Phys. {\bf 93} (1995), 757. 
\bibitem{rf:Mitsumori}
T.~Mitsumori, N.~Noda, K.~Koide, H.~Kouno A. Hasegawa and M.~Nakano, 
Prog. Theor. Phys. {\bf 96} (1996), 179. 
\bibitem{rf:Noda}
N.~Noda, A.~Hasegawa, H.~Kouno and M.~Nakano, 
Prog. Theor. Phys. {\bf 97} (1997), 451. 
\bibitem{rf:Pearson}
J.~M.~Pearson, Phys. Lett. {\bf B271}~(1991), 12. 
\end{thebibliography}
\end{document}